\RequirePackage{scrlfile}
\AfterClass{article}{\RequirePackage{float}}

\documentclass[3p]{elsarticle}

\usepackage[hyphens]{url}  
\usepackage{jltdefs}
\usepackage{stmaryrd}  
\graphicspath{{figs/}{figs_unused/}}
\usepackage{tikz}
\usepackage{float}

\def\appendixname{}

\newcommand{\overbar}[1]{
  \mkern 1.5mu\overline{\mkern-1.5mu#1\mkern-1.5mu}\mkern 1.5mu}

\newcommand{\keep}[1]{}

\mathnotation{\hf}{\hat{f}}
\mathnotation{\htheta}{\skew4\hat{\theta}}
\mathnotation{\hTs}{{{}\widehat{S}}}
\mathnotation{\hT}{{{}\widehat{\T}}}
\mathnotation{\hU}{{{}\widehat{\U}}}

\renewcommand{\t}{t}
\mathnotation{\tcook}{\t_{\mathrm{cook}}} 
\mathnotation{\tmincook}{\t_{\mathrm{mincook}}} 

\mathnotation{\s}{s}
\mathnotation{\z}{z}
\mathnotation{\h}{h}
\mathnotation{\Ll}{L}
\mathnotation{\T}{T}
\mathnotation{\Th}{\T_{\mathrm{H}}}
\mathnotation{\Tc}{\T_{\mathrm{C}}}
\mathnotation{\Ts}{S}
\mathnotation{\dT}{\Delta\T}
\mathnotation{\Tcook}{\T_{\mathrm{cook}}} 
\mathnotation{\eigf}{\phi}   
\mathnotation{\eigv}{\mu}    
\mathnotation{\U}{U}
\mathnotation{\Ub}{\overbar{\U}}
\mathnotation{\dU}{\Delta\U}

\mathnotation{\cF}{\mathcal{F}}
\mathnotation{\cG}{\mathcal{G}}
\mathnotation{\Tf}{\cF\T}
\mathnotation{\Tsf}{\cF\Ts}
\mathnotation{\eigff}{\cF\eigf}

\mathnotation{\cH}{\mathcal{H}}
\mathnotation{\cK}{\mathcal{K}}
\mathnotation{\dt}{\Delta\t}
\mathnotation{\dz}{\Delta\z}
\mathnotation{\eye}{I}

\mathnotation{\Nflips}{N}
\mathnotation{\tcond}{k}                 
\mathnotation{\flipp}{F}

\mathnotation{\wT}{\widetilde\T}
\mathnotation{\wz}{\tilde\z}
\mathnotation{\wt}{\tilde\t}
\mathnotation{\wh}{\tilde\h}

\mathnotation{\dTs}{\Delta\Ts}
\mathnotation{\tct}{\t_{\text{cookthrough}}}

\mathnotation{\degreeC}{{}^\circ\mathrm{C}}
\mathnotation{\centimeter}{\mathrm{cm}}
\mathnotation{\meter}{\mathrm{m}}
\mathnotation{\kilogram}{\mathrm{kg}}
\ifdefined\second 
\renewcommand{\second}{\mathrm{s}}
\else
\mathnotation{\second}{\mathrm{s}}
\fi
\mathnotation{\watt}{\mathrm{W}}
\mathnotation{\joule}{\mathrm{J}}

\newcommand{\matlabfootnote}[1]{\footnote{#1}}
\newcommand{\MATLAB}{MATLAB}
\newcommand{\giturl}{https://github.com/jeanluct/cookflip_code}
\newcommand{\blob}{/blob/main}
\newcommand{\matlabcommand}[1]{%
  {\href{\giturl\blob/#1.m}{\texttt{#1}}}}

\begin{document}

\title{
  The mathematics of burger flipping}

\author{Jean-Luc Thiffeault}
\affiliation{Department of Mathematics, University of Wisconsin --
  Madison, 480 Lincoln Dr., Madison, WI 53706, USA}
\ead{jeanluc@math.wisc.edu}

\date{}

\begin{abstract}
  What is the most effective way to grill food?  Timing is everything, since
  only one surface is exposed to heat at a given time.  Should we flip only
  once, or many times?  We present a simple model of cooking by flipping, and
  some interesting observations emerge.  The rate of cooking depends on the
  spectrum of a linear operator, and on the fixed point of a map.  If the
  system has symmetric thermal properties, the rate of cooking becomes
  independent of the sequence of flips, as long as the last point to be cooked
  is the midpoint.  After numerical optimization, the flipping intervals
  become roughly equal in duration as their number is increased, though the
  final interval is significantly longer.  We find that the optimal
  improvement in cooking time, given an arbitrary number of flips, is
  about~$29\%$ over a single flip.  This toy problem has some characteristics
  reminiscent of turbulent thermal convection, such as a uniform average
  interior temperature with boundary layers.
\end{abstract}

\maketitle


\medskip

\noindent
\textsl{Dedicated to Charlie Doering, a man who appreciated a good burger and
  good company.}

\section{Introduction}
\label{sec:intro}

What is the best strategy for grilling a steak, a burger, or a slice of
eggplant on a hot grill?  Only one side is hot at a time, so timing is
crucial.  One school of thought is that flipping only once leads to more even
cooking.  Kenji L\'{o}pez-Alt from The Food Lab reports on this
\cite{FoodLab_flipping}:
\begin{quote}
  As food scientist and writer Harold McGee has pointed out, flipping steak
  repeatedly during cooking can result in a cooking time about 30\% faster
  than flipping only once. The idea is that with repeated flips, each surface
  of the meat is exposed to heat relatively evenly, with very little time for
  it to cool down as it faces upwards. The faster you flip, the closer your
  setup comes to approximating a cooking device that would sear the meat from
  both sides simultaneously.
\end{quote}
This raises the intriguing possibility of making a mathematical model that
demonstrates this potential faster cooking.  Though our model will be very
simple, its simplicity will allow us to isolate the mathematical elements that
induce faster cooking, opening the way to optimization.

As the simplest possible model, we take the food to be an infinite slab and
only keep track of its temperature in the vertical direction.  We solve the
standard heat equation in that slab.  Though we do not include more
complicated aspects such as moisture and fat content, we do allow for
imperfect conductor boundary conditions at the top and bottom of the food.
Once we understand this very basic problem, we allow for a flip of the food
--- that is, we turn over the food on the heating surface.  Mathematically, we
overturn the vertical temperature distribution and allow the heat equation to
act, using the flipped profile as an initial condition.  We can express the
flipping-and-heating operation as a map acting on the Fourier coefficients of
the solution.  We will see that understanding the fixed point and the spectral
properties of this map is crucial to optimizing the total cooking time.

The outline of the paper is as follows.  In \cref{sec:model} we introduce the
model system and discuss relevant physical parameter values and our
dimensionless scalings.  In~\cref{sec:solving} we provide a solution to the
simple one-dimensional model, in the form of a sum over separated solution
using Sturm--Liouville theory.  This is elementary, but we briefly give the
details for the sake of completeness.  \cref{sec:noflip} is devoted to the
simplest cooking problem: how long does it take for the food to cook if it is
never flipped?  Depending on the boundary conditions, the food might never
actually cook in this configuration.  In \cref{sec:flipping} we finally add
nontrivial dynamics: we flip the food at prescribed periodic times and examine
the interior temperature.  We define the \emph{flip-heat operator}, which is a
composition of a flip followed by a cooking interval.  \Cref{sec:sym} treats
the symmetric limit where the food has identical thermal properties at the top
and bottom, which allows for further analytic progress (this is the Boussinesq
limit in thermal convection~\cite{Zhang1997}).  We answer the burning question
of optimal cooking times in \cref{sec:opt}: what is the optimal sequence of
flips to achieve fastest cooking?  Most of the results of this section are
numerical; the evidence suggests that the problem has a unique global minimum.
We offer some concluding remarks in \cref{sec:discussion}.%
\matlabfootnote{Throughout the paper, footnotes indicate when a particular
  \MATLAB\ script can be used to reproduce a computation.  The code can be
  downloaded on \url{\giturl}.}

\section{The model equations and physical parameters}
\label{sec:model}


We consider a piece of food that is relatively flat and thin, being cooked on
a heating surface, and that can be turned over --- flipped --- to assist
cooking.  We model the food as an infinite slab, extending from~$\wz=0$
to~$\wz=\Ll$ in the vertical (see \cref{fig:slab}).  In dimensional form, the
temperature~$\wT(\wz,\wt)$ in the food satisfies the heat equation
\begin{subequations}
\begin{equation}
  \wT_{\wt} = \kappa\wT_{\wz\wz},\qquad
  0 < \wz < \Ll,\quad  \wt > 0,
\end{equation}
with initial condition
\begin{equation}
  \wT(\wz,0) = \wT_0(\wz).
\end{equation}
(The subscripts~$\wt$ and~$\wz$ denote partial derivatives.)  Newton's law of
cooling gives the boundary conditions~\cite[p.~7]{Pinsky}
\begin{equation}
  -\tcond\,\wT_{\wz}(0,\wt) = \wh_0\,(\Th - \wT(0,\wt)),
  \qquad
  \tcond\,\wT_{\wz}(\Ll,\wt) = \wh_1\,(\Tc - \wT(\Ll,\wt)),
\end{equation}
\label{eq:tilde}%
\end{subequations}
where~$\wh_0>0$ and~$\wh_1\ge0$, in order that heat flow obeys the second law
of thermodynamics.  For~$\wh_i\rightarrow\infty$ we recover a perfect
conductor (fixed-temperature boundary condition), and for~$\wh_1=0$ we have a
perfect insulator, which we disallow at the heating surface ($\wh_0>0$).

\begin{figure}
\begin{center}
  \begin{tikzpicture}[scale=3]
    \fill[lightgray] (0,0) rectangle (2,-.25);
    \fill[lightgray] (0,1) rectangle (2,1.25);
    \draw[very thick,-]
    (0,0) node[left] {$\wz=0$} -- (1,0) node[below] {$\Th$ (metal)} -- (2,0)
    node[right] {bottom};
    \draw[very thick,-]
    (0,1) node[left] {$\wz=\Ll$} -- (1,1) node[above] {$\Tc$ (air)} -- (2,1)
    node[right] {top};
    \draw[very thick,-,red]
    (.2,1) to[bend left]
    (1,.5)
    node[right,black] {\hspace*{.5em}$\wT(\wz,\wt)$}
    node[left,black] {food interior\hspace*{1em}}
    to[bend right] (1.8,0);
  \end{tikzpicture}
\end{center}
\caption{Slab geometry for the food.  In general $\wT(0,\wt)<\Th$ and
  $\wT(\Ll,\wt)>\Tc$, with equality only for perfect conductors.}
\label{fig:slab}
\end{figure}

The physical constants involved can vary wildly for different types of food
and cooking surfaces, and for simplicity we will mostly confine ourselves to
one set of parameters as listed in \cref{tab:physical}, unless otherwise
noted.  For a meat patty, we get physical parameters from \citet{Zorrilla2003}
(see also~\cite{Stroshine}).  At the bottom, they use a `contact heat transfer
coefficient' $\h_0 = 900\,\watt/\meter^2\,\degreeC$ at the interface of the
heating surface and the meat ($\wz=0$) \cite[p.~62]{Zorrilla2003}, which lumps
together the resistance of a thin layer of fat, air, and moisture.  At the
top, they take a combined radiation and convection heat transfer coefficient
$\h_1 = 60\,\watt/\meter^2\,\degreeC$.  In the interior of the food, they use
a thermal conductivity $\tcond = 0.416\,\watt/\meter\,\degreeC$.  The
volumetric heat capacity is
$\rho c = 3.4533\times 10^6\,\joule/\meter^3\,\degreeC$, so the thermal
diffusivity is
$\kappa = \tcond/\rho c = 1.205\times 10^{-7}\,\meter^2/\second$.  (For
comparison, water is about~$1.4 \times 10^{-7}\,\meter^2/\second$.)  We will
take the thickness of the food to be~$\Ll = 10^{-2}\,\meter$.  A hot plate is
typically at about~$\Th = 200\,\degreeC$, the ambient air temperature
about~$\Tc = 25\,\degreeC$, so~$\dT = 175\,\degreeC$.  A good temperature for
cooked beef is about~$\Tcook = 70\,\degreeC$.

Given all these dimensional parameters, we nondimensionalize the system by
using~$\Ll = 10^{-2}\,\meter$ as a length scale.  For a time scale we use the
typical conductive time scale~$\Ll^2/\kappa = 830\,\second$ (almost~$14$
minutes).  The temperature scale is~$\dT = \Th-\Tc = 175\,\degreeC$.
Finally, we fix the energy scale to
be~$\tcond\dT\Ll^3/\kappa = 604.3\,\joule$.  The physical parameters and their
nondimensional versions are summarized in \cref{tab:physical}.


\begin{table}[H]
  \caption{Physical parameters used in the paper.  When made dimensionless,
    temperatures are shifted by~$\Tc$ before being scaled by~$\dT$.}
\label{tab:physical}
\begin{center}
\begin{tabular}{cccl}
\hline\hline
notation & value & d'less & description \\[2pt]
\hline
$\Ll$ & $0.01\,\meter$ & $1$ & food thickness (length scale) \\
$\Ll^2/\kappa$ & $830\,\second$ & $1$ & diffusive time (time scale) \\
$\dT$ & $175\,\degreeC$ & $1$ & temperature difference (temperature scale) \\
$\Th$ & $200\,\degreeC$ & $1$ & bottom temperature \\
$\Tc$ & $25\,\degreeC$ & $0$ & top temperature (also initial temp.) \\
$\Tcook$ & $70\,\degreeC$ & $0.257$ & cooked temperature \\
$\h_0$ & $900\,\watt/\meter^2\,\degreeC$ & $21.6$ & bottom heat transfer
coefficient \\
$\h_1$ & $60\,\watt/\meter^2\,\degreeC$ & $1.44$ & top heat transfer
coefficient \\
$\kappa$ & $1.205\times 10^{-7}\,\meter^2/\second$ & $1$ & thermal
diffusivity of food \\
\tcond & $0.416\,\watt/\meter\,\degreeC$ & $1$ & thermal conductivity of
food \\
$\rho c$ & $3.4533\times 10^6\,\joule/\meter^3\,\degreeC$ & $1$ & volumetric
heat capacity of food \\
$\eigv_1$ & & 2.0803 & first eigenvalue for \cref{eq:trans} \\
$\eigv_2$ & & 4.7865 & second eigenvalue \\
$\eigv_3$ & & 7.6966 & third eigenvalue \\
\hline\hline
\end{tabular}
\end{center}
\end{table}


We make the system dimensionless by letting~$\z = \wz/\Ll$,
$\t = \kappa\wt/\Ll^2$, and~$\T(\z,\t) = (\wT(\wz,\wt) - \Tc)/\dT$
with~$\dT=\Th-\Tc$; we then have the PDE system
\begin{subequations}
\begin{alignat}{3}
  \T_\t &= \T_{\z\z},\qquad &0 < \z < 1,&\quad  &\t > 0, \\
  \T(\z,0) &= \T_0(\z),\qquad &0 < \z < 1,& \\
  \T_\z(0,\t) &= -\h_0\,(1 - \T(0,\t)), & & &\t > 0,
  \label{eq:BCT0} \\
  \T_\z(1,\t) &= -\h_1\,\T(1,\t), & & &\t > 0,
  \label{eq:BCT1}
\end{alignat}
\label{eq:inhomog}%
\end{subequations}
where~$\h_i = \Ll\wh_i/\tcond$.  In the remainder of the paper we use
dimensionless quantities unless otherwise noted.  We will typically take
$\T_0(\z) \equiv 0$: the food starts at room temperature.%
\matlabfootnote{The \MATLAB\ function \matlabcommand{heat}
  solves \cref{eq:inhomog} for a given initial condition.}

\section{Solving the heat equation}
\label{sec:solving}

The solution of the system~\eqref{eq:inhomog} is a standard exercise in linear
PDEs \cite{Pinsky}.  We briefly outline the method here for completeness, and
to fix the notation.

\subsection{The steady profile and temperature deviation}
\label{sec:steady}

Let's solve the steady problem for \cref{eq:inhomog} first, with the steady
linear profile~$\T(\z,\t) = \Ts(\z) = a + b\z$.  After applying the boundary
conditions~\eqref{eq:BCT0}--\eqref{eq:BCT1}, we find%
\matlabfootnote{The \MATLAB\ function \matlabcommand{heatsteady} computes the
  steady temperature profile~\eqref{eq:Ts}.}
\begin{equation}
  \Ts(\z) = \frac{\h_0(1 + \h_1 - \h_1\z)}{\h_0 + \h_1 + \h_0\h_1}
  =
  \Ts(0) - \dTs\,\z,
  \qquad
  \dTs \ldef \Ts(0) - \Ts(1),
  \label{eq:Ts}
\end{equation}
with
\begin{equation}
  \Ts(0) = \l(1 + \frac{\h_1/\h_0}{1+\h_1}\r)^{-1} \le 1, \qquad
  \Ts(1) = \l(1 + \h_1 + \frac{\h_1}{\h_0}\r)^{-1} \le 1.
\end{equation}
For large~$\h_0$ and small~$\h_1$, we have
\begin{equation}
  \Ts(0) \simeq 1 - \h_1/\h_0, \qquad
  \Ts(1) \simeq 1 - \h_1,
\end{equation}
so both the top and bottom temperatures are near~$1$, that is, the steady
profile is nearly uniform.  For~$\h_0=\infty$ and~$\h_1=0$, we
have~$\Ts(\z) = 1$.

We use \cref{eq:Ts} to reformulate~\eqref{eq:inhomog} as a homogeneous problem
for the temperature deviation
\begin{equation}
  \theta(\z,\t) = \T(\z,\t)-\Ts(\z),
\end{equation}
which satisfies
\begin{subequations}
\begin{alignat}{3}
  \theta_\t &= \theta_{\z\z},\qquad &0 < \z < 1,&\quad  &\t > 0, \\
  \theta(\z,0) &= \theta_0(\z) = \T_0(\z)-\Ts(\z),\qquad &0 < \z < 1,& \\
  \theta_\z(0,\t) &= \h_0\,\theta(0,\t), & & &\t > 0,
  \label{eq:BCtheta0} \\
  \theta_\z(1,\t) &= -\h_1\,\theta(1,\t), & & &\t > 0.
  \label{eq:BCtheta1}
\end{alignat}
\label{eq:homog}%
\end{subequations}
The typical initial condition $\T_0(\z) \equiv 0$ (room temperature) becomes
$\theta_0(\z) = -\Ts(\z)$.  In \cref{sec:pdesol} we show how to solve
\cref{eq:homog}, and hence \cref{eq:inhomog,eq:tilde}, by an eigenfunction
expansion.

\subsection{Solution by eigenfunction expansion}
\label{sec:pdesol}

The PDE system~\eqref{eq:homog} is solved by separation of variables.  Writing
the separated solution~$\theta(\z,\t)=\eigf(\z)\tau(\t)$, we have
\begin{equation}
  \frac{\tau_\t}{\tau} = \frac{\eigf_{\z\z}}{\eigf} = -\eigv^2 =
  \text{const.},
  \label{eq:sep}
\end{equation}
where~$\eigv^2 \in \mathbb{R}$ is the separation constant.  For~$\eigv=0$, we
have~$\tau = \text{const.}$ and~$\eigf(\z) = A + B\z$.  The boundary
conditions~\eqref{eq:BCtheta0}--\eqref{eq:BCtheta1} lead
to~$(\h_0+\h_1+\h_0\h_1)A=0$, so we get the trivial solution; hence, we must
have~$\eigv \ne 0$.  It is straightforward to show that imaginary $\eigv$ can
be ruled out by the boundary conditions as well~\cite{Pinsky}.

Taking $\eigv>0$ without loss of generality, we write the general separated
solution to \cref{eq:sep} as
\begin{equation}
  \tau(\t) = \ee^{-\eigv^2\t},\qquad
  \eigf(\z) = A\,\cos\eigv\z + B\,\sin\eigv\z.
\end{equation}
Applying the boundary conditions \cref{eq:BCtheta0,eq:BCtheta1}
to~$\eigf(\z)$, we obtain~$A\h_0=\eigv B$ and the transcendental equation
\begin{equation}
  \frac{(\h_0 + \h_1)\eigv}{\eigv^2 - \h_0\h_1} = \tan\eigv,
  \qquad
  \eigv > 0.
  \label{eq:trans}
\end{equation}

In general the possible solutions~$\eigv_m$, $m=1,2,3,\dots$, have to be found
numerically.%
\matlabfootnote{See \MATLAB\ functions \matlabcommand{heateigval} and
  \matlabcommand{heateigfun}.}
We order them such that~$\eigv_1 < \eigv_2 < \dots$.  Note that these are all
distinct, being the eigenvalues of a Sturm--Liouville
problem~\cite[p.~86]{Pinsky}.
\keep{%
  \Cref{fig:trans} shows the left-hand and right-hand sides of \cref{eq:trans}
  for the parameters in \cref{tab:physical}.
  \begin{figure}
    \begin{center}
      \includegraphics[width=.6\textwidth]{trans}
    \end{center}
    \caption{Left-hand side (dashed blue) and right-hand side (solid red) of
      \cref{eq:trans} for~$\h_0=21.6$, $\h_1=1.44$.  The intersections
      correspond to eigenvalues.}
    \label{fig:trans}
  \end{figure}
}
For the parameters in \cref{tab:physical}, the first few roots are
$\eigv_1\approx2.0803$, $\eigv_2\approx4.7865$, $\eigv_3\approx7.6966$,
$\eigv_4\approx10.6709$.  The root~$\eigv_m$ approaches a multiple of~$\pi$
for $m$ large.  \keep{$m\pi$ or $(m-1)\pi$.}

The~$\mathrm{L}^2$-normalized eigenfunctions corresponding to the~$\eigv_m$
can be written
\begin{equation}
  \eigf_m(\z) = C_m^{-1}\l(\sin\eigv_m\z + (\eigv_m/\h_0)\cos\eigv_m\z\r),
  \qquad m = 1,2,3,\dots.
  \label{eq:eigenf}
\end{equation}
with the normalization constant~$C_m>0$ defined as
\begin{equation}
  C_m^2
  = \tfrac12(1 + (\eigv_m/\h_0)^2)
  + \h_0^{-1}\sin^2\eigv_m
  + \tfrac14(\eigv_m\h_0^{-2} - \eigv_m^{-1})\sin2\eigv_m.
  \label{eq:Cm2}
\end{equation}
Because they are Sturm--Liouville eigenfunctions, the~$\eigf_m$ satisfy
orthogonality relations with respect to the standard~$\mathrm{L}^2$ inner
product:
\begin{equation}
  \langle\eigf_m,\eigf_n\rangle =
  \delta_{mn},
  \qquad
  \langle f,g\rangle \ldef \int_0^1 f(\z)\,g(\z)\dint\z.
  \label{eq:inner}
\end{equation}

Now we expand the initial condition for~$\theta(\z,\t)$ in terms of these
orthogonal eigenfunctions to obtain a generalized Fourier series%
\footnote{Generalized in the sense that the orthogonal functions are
  the~$\eigf_m$s, not the individual sines and cosines.}
\begin{equation}
  \theta_0(\z) = \sum_{m=1}^\infty {\htheta_0}_m\eigf_m(\z),
\end{equation}
where for any function~$f(\z)$ the Fourier coefficients are
\begin{equation}
  {\hf}_m = \langle f,\eigf_m\rangle.
\end{equation}
Given this, the full solution to~$\theta(\z,\t)$ in~\eqref{eq:homog} is
immediate:
\begin{equation}
  \theta(\z,\t)
  =
  \cH_\t\theta_0(\z)
  =
  \sum_{m=1}^\infty
  {\htheta_0}_m\,\ee^{-\eigv_m^2\t}\,\eigf_m(\z).
  \label{eq:thetasol}
\end{equation}
Here we define the heat operator~$\cH_\t$ and heat kernel~$H$ as
\begin{equation}
  \cH_\t \theta_0(\z)
  =
  \int_0^1 H(\z,\t;\z',0) \,\theta_0(\z')\dint\z',
  \quad
  H(\z,\t;\z',0)
  =
  \sum_{m=1}^\infty
  \ee^{-\eigv_m^2\t}
  \eigf_m(\z)\,\eigf_m(\z').
  \label{eq:heatopkernel}
\end{equation}
We add the steady profile~$\Ts(\z)$ to~\eqref{eq:thetasol} to obtain the
solution to~\eqref{eq:inhomog}:
\begin{equation}
  \T(\z,\t)
  =
  \Ts(\z)
  +
  \cH_\t\theta_0(\z)
  =
  \Ts(\z)
  +
  \sum_{m=1}^\infty {\htheta_0}_m\,\ee^{-\eigv_m^2\t}\,\eigf_m(\z).
  \label{eq:Tsol3}
\end{equation}
We can expand~$\Ts(\z)$ itself in terms of the eigenfunctions:
\begin{equation}
  \Ts(\z) = \sum_{m=1}^\infty {\hTs}_m\,\eigf_m(\z),
\end{equation}
so our typical initial condition $\theta_0(\z) = -\Ts(\z)$ becomes
${\htheta_0}_m = -{\hTs}_m$ in terms of Fourier coefficients.  Since the
Fourier coefficients of the steady profile~$\Ts(\z)$ play an important role,
we observe that we can compute them directly by integrating \cref{eq:Ts}
against~\eqref{eq:eigenf} to obtain
\begin{equation}
  {\hTs}_m
  =
  (\eigv_m\,C_m)^{-1} > 0
  \label{eq:hTsm}
\end{equation}
where~$C_m$ is defined in \cref{eq:Cm2}.


\keep{I don't think this is useful? %
\begin{align*}
 \eigv_m^2C_m{\hTs}_m
 &=
 \int_0^1 (\Ts(0) - \dTs\,\z)\,\eigv_m^2\eigf_m(\z)\dint\z \nonumber\\
 &=
 -\int_0^1 (\Ts(0) - \dTs\,\z)\,{\eigf_m}_{\z\z}(\z)\dint\z \nonumber\\
 &=
 -\Ts(0)\,\l[{\eigf_m}_\z(\z)\r]_0^1
 + \dTs
 \l(\l[\z\,{\eigf_m}_\z(\z)\r]_0^1
 - \int_0^1 {\eigf_m}_\z(\z)\dint\z\r) \nonumber\\
 &=
 -\Ts(0)\,({\eigf_m}_\z(1) - {\eigf_m}_\z(0))
 + \dTs
 \l({\eigf_m}_\z(1)
 - \eigf_m(1) + \eigf_m(0)\r) \nonumber\\
 &=
 \Ts(0)\,(\h_1\eigf_m(1) + \h_0\eigf_m(0))
 - \dTs
 \l(\h_1\eigf_m(1)
 + \eigf_m(1) - \eigf_m(0)\r) \nonumber\\
 &=
 \h_0\, \Ts(0)\,\eigf_m(0)
 + \h_1\, \Ts(1)\,\eigf_m(1)
 +
 \dTs\,(\eigf_m(0) - \eigf_m(1))
\end{align*}
}

\section{Cooking without flipping}
\label{sec:noflip}

Let's examine what happens if we just leave the food on the heating plate for
a long time, without flipping.  The Fourier coefficients of~$\theta(\z,\t)$
evolve according to \cref{eq:thetasol},
\begin{equation}
  {\htheta}_m(\t) = -\hTs_m\,\ee^{-\eigv_m^2\t}
\end{equation}
where we took~${\htheta_0}_m = -\hTs_m$ (\ie, $\T_0 \equiv 0$).  For
moderately large~$\t$, let's approximate the temperature at time~$\t$ by
keeping only the first eigenfunction:
\begin{equation}
  \T(\z,\t) \approx \Ts(\z)
  - \hTs_1\,\ee^{-\eigv_1^2\t}\eigf_1(\z).
  \label{eq:T1mode}
\end{equation}
Now assume we have a dimensionless target cooking temperature~$\Tcook$,
with~$0 < \Tcook \le \Ts(0)$, where~$\Ts(0)$ is the hottest point of the
steady solution.  The criterion for the food to have cooked through to~$\z=1$
without needing to flip is
\begin{equation}
  \Ts(1) - \hTs_1\,\ee^{-\eigv_1^2\t}\,\eigf_1(1) = \Tcook,
  \label{eq:tcookthrough_crit}
\end{equation}
where~$\Ts(1)$ is the coldest point of the steady solution.  We can
solve~\eqref{eq:tcookthrough_crit} for~$\t$ to obtain the `cookthrough time'
for the food to cook without needing a flip:
\begin{equation}
  \tct = \frac{1}{\eigv_1^2}\log\l(
  \frac{\hTs_1\eigf_1(1)}{\Ts(1) - \Tcook}\r),
  \qquad
  \ee^{-(\eigv_2^2 - \eigv_1^2)\,\tct} \ll 1.
  \label{eq:tcookthrough}
\end{equation}
Here~$\hTs_1\eigf_1(1) > 0$ for $\h_1 < \infty$, and we need~$\Tcook < \Ts(1)$
for the cookthrough time to be well-defined.  We conclude that for
$\Ts(1) \le \Tcook \le \Ts(0)$, flipping the food is required for cooking.  In
particular, for $\h_1 = \infty$ (perfect conductor at the top), we have
$\eigf_1(1)=0$, and the top of the food remains at zero temperature no matter
how long we wait.

\keep{We have $\hTs_m > 0$ from \cref{eq:hTsm}, and
  \begin{equation}
    \eigf_m(1)
    =
    \frac{1}{C_m}\,\frac{\h_0^2 + \eigv_m^2}{\h_0(\h_0 + \h_1)}
    \,\sin\eigv_m\,.
  \end{equation}
  From \cref{fig:trans} it is clear that $\sgn(\sin\eigv_m) = (-1)^{m+1}$, so
  $\eigf_1(1) > 0$ for $\h_1 < \infty$.

  \begin{equation}
    \Ts(1) =
    \sum_{m=1}^\infty \hTs_m \eigf_m(1)
  \end{equation}
}

For the parameters in \cref{tab:physical}, we have $\tct \approx 0.340$, or
$283\,\second$ in dimensional terms.%
\matlabfootnote{See \MATLAB\ function \matlabcommand{tcookthru}; for our
  parameters, the approximate solution \cref{eq:tcookthrough} is accurate to
  $0.07\%$.}
This is not that long, since the food is fairly thin, but we will see that the
cooking time can be greatly shortened by flipping, as of course experience
suggests.

\section{Flipping the food}
\label{sec:flipping}

In \cref{sec:solving} we solved the PDE \eqref{eq:inhomog} and derived the
time evolution of~$\T(\z,\t)$ in terms of Fourier coefficients, as given by
\cref{eq:Tsol3}.  We then found in \cref{sec:noflip} that it is sometimes
possible to cook without flipping, as long as the steady top temperature
$\Ts(1) > \Tcook$, though this can take a long time depending on the boundary
conditions.  In the present section we model a `flip' of the food, that is,
turning the food over on the hot plate.

\subsection{Flipping at fixed intervals}
\label{sec:flipfixed}

There are two possible, mathematically-equivalent, approaches to deal with the
`flipping' of the food on the hot surface: flip the food (the~$[0,1]$ domain
itself), or flip the boundary conditions.  In the context of solid and fluid
mechanics, we can think of these as the Eulerian (fixed frame) and Lagrangian
(moving frame) pictures, respectively.  Here it is fairly easy to convert
between both pictures, and we choose to flip the food (Eulerian).  With this
choice we have to ensure the $\zc$ coordinate labels the same point in the
food when determining if it is cooked.  We will return to this issue in
\cref{sec:opt}, when we look for optimal solutions.

Consider a slab with vertical temperature distribution~$\T(\z)$, $0<\z<1$.
If we flip the slab over, we obtain its new vertical temperature by
replacing~$\z$ by~$1-\z$.  We write this in terms of the `flipping operator'
defined by
\begin{equation}
  \cF f(\z) = f(1-\z).
  \label{eq:flip}
\end{equation}
It is easy to see that this operator is self-adjoint with respect to the inner
product~\eqref{eq:inner}.  It has eigenvalues $\pm1$ and its eigenfunctions
consist of even and odd functions with respect to~$\z=1/2$.

Now define the `flip-and-heat' operation, where we take an initial heat
profile~$\T(\z,\t)$, flip it using~\eqref{eq:flip}, then allow it to evolve
for a time~$\dt$.\footnote{Since our initial condition~$\T(\zc,0)=0$
  is~$\cF$-invariant, it is immaterial whether we flip at the beginning or the
  end of the interval~$[\t,\t+\dt]$.}  From \cref{eq:Tsol3}, the temperature
profile at time~$\t + \dt$ is
\begin{equation}
  \T(\z,\t+\dt)
  =
  \Ts(\z)
  +
  \cH_{\dt}(\cF\T(\z,\t) - \Ts(\z)).
  \label{eq:flipheatmap}
\end{equation}
We can solve this recurrence relation to obtain at time~$\t_k = k\dt$
\begin{align}
  \T(\z,\t_k)
  &=
  \cK_{\dt}^k\,
  \T_0(\z)
  +
  \sum_{j=0}^{k-1}
  \cK_{\dt}^j\,(1 - \cH_{\dt})\Ts(\z)
  \nonumber \\
  &=
  \cK_{\dt}^k\,
  \T_0(\z)
  +
  (1 - \cK_{\dt})^{-1}
  (1 - \cK_{\dt}^k)\,(1 - \cH_{\dt})\Ts(\z)
  \label{eq:recsoleqdt}
\end{align}
where we used the geometric sum formula and defined the \emph{flip-heat
  operator} as
\begin{equation}
  \cK_{\dt} \ldef \cH_{\dt}\cF.
  \label{eq:flipheat}
\end{equation}
Since~$\lVert\cK_{\dt}\rVert < 1$, \keep{(see \cref{sec:Kdtless1}),}
$\T(\z,\t_k)$ converges to
\begin{equation}
  \U_{\dt}(\z)
  \ldef
  (1 - \cK_{\dt})^{-1}\,(1 - \cH_{\dt})\Ts(\z)
  \label{eq:U}
\end{equation}
as~$k \rightarrow
\infty$.  This is the fixed point for the map~\eqref{eq:flipheatmap}.  The
rate of convergence is determined by the modulus of the eigenvalues
of~$\cK_{\dt}$.  Note that in a material (Lagrangian) frame moving with the
food,~$\U_{\dt}(\z)$ is \emph{not} a fixed point, but rather it flips at every
interval.

\begin{figure}
  \begin{center}
    \subfigure[]{
      \includegraphics[width=.46\textwidth]{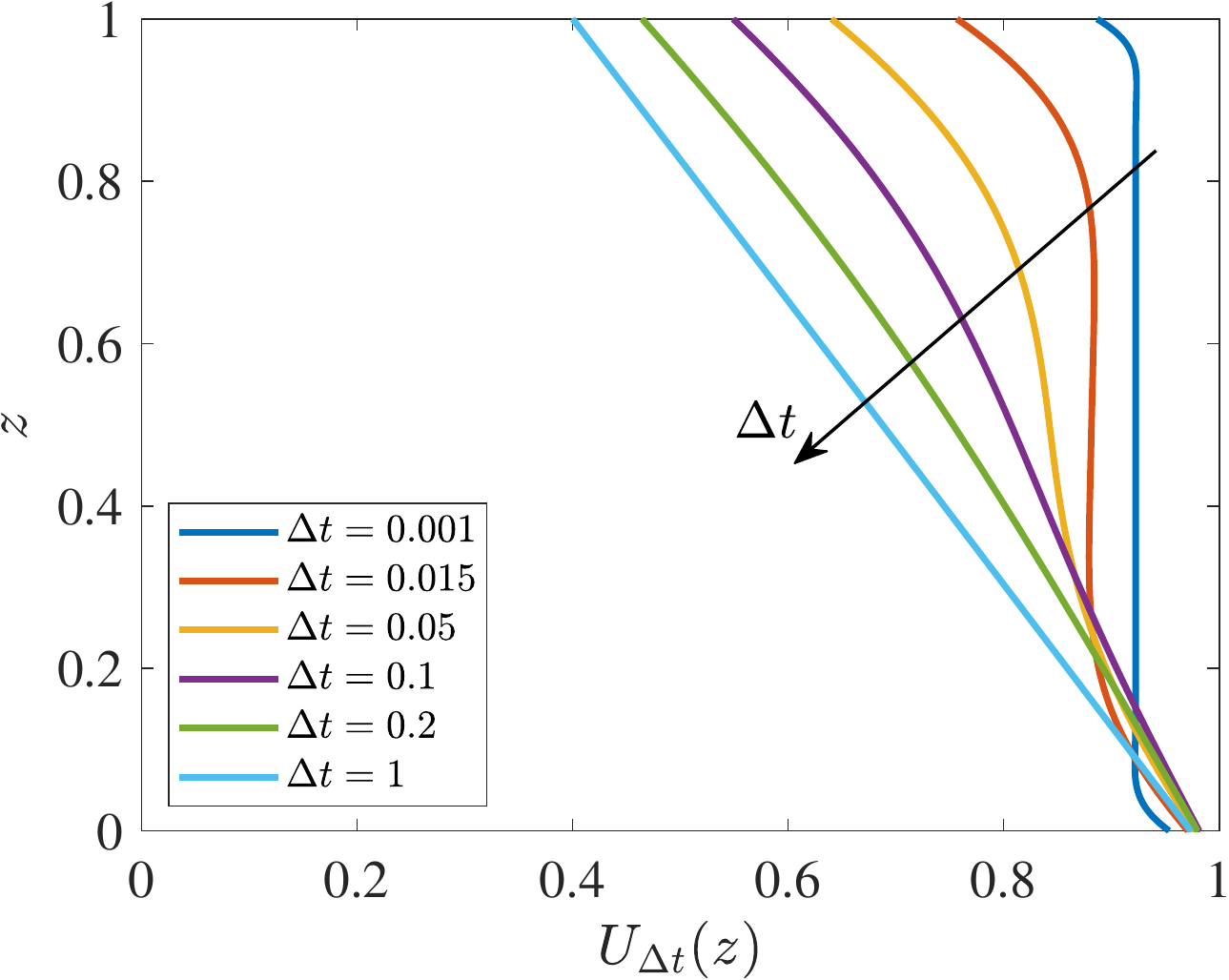}
      \label{fig:flipheatfix}
    }\hspace{.02\textwidth}%
    \subfigure[]{
      \includegraphics[width=.46\textwidth]{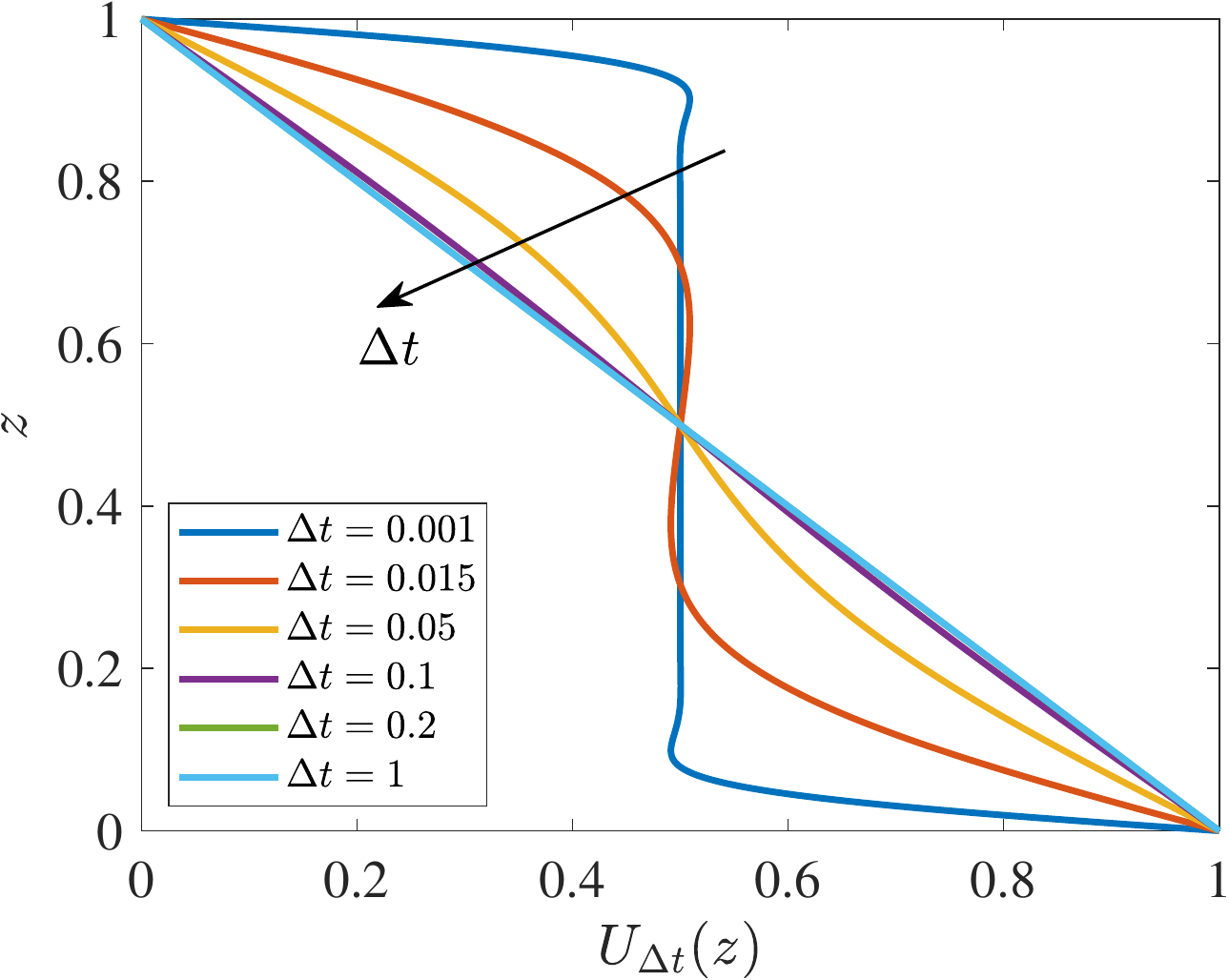}
      \label{fig:flipheatfix_sym}
    }
  \end{center}
  \caption{Asymptotic temperature profile $\U_{\dt}(\z)$ for (a) the physical
    parameters from \cref{tab:physical}; (b) perfect conductors
    $\h_0=\h_1=\infty$ at top and bottom.  The better insulation at the top in
    (a) leads to much higher temperatures.  In both cases, rapid flipping
    leads to a uniform asymptotic temperature~$\Ub$ in the interior, with
    boundary layers near the edges.}
  \label{fig:flipheatfix_both}
\end{figure}

\Cref{fig:flipheatfix_both} shows numerical solutions for $\U_{\dt}(\z)$ for
several values of~$\dt$, for the reference values in \cref{tab:physical}
(\cref{fig:flipheatfix}) and for fixed temperature at both boundaries
(\cref{fig:flipheatfix_sym}, $\h_0=\h_1=\infty$).%
\matlabfootnote{See \MATLAB\ function \matlabcommand{flipheatfix}.}
For large $\dt$, $\U_{\dt}$ converges to the steady conduction profile $\Ts$.
For small $\dt$, $\U_{\dt}$ limits to a constant temperature in the interior,
with boundary layers at~$\z=0$ and~$1$.  The presence of the boundary layers
can be attributed to the rapid flipping: the heat flux from the edges only
penetrates a depth $\sqrt{\dt}$ in the interior at each flip.  This kind of
profile is reminiscent of the mean temperature in turbulent
Rayleigh--B\'{e}nard convection~\cite[Fig.~15]{Hansen1992}.  The interior is
`well-mixed,' so the temperature is uniform, with boundary layers in order to
satisfy the boundary conditions.  The small overshoot seen in
\cref{fig:flipheatfix_sym} is also present in convection, though the overshoot
there tends to be larger.  (In \cref{sec:blayer} we derive a precise
expression for the boundary layer solution in \cref{fig:flipheatfix_sym} for
the limit of small~$\dt$.)

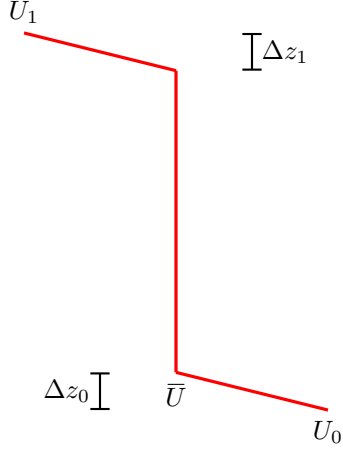
\begin{figure}[t]
  \begin{center}
    \begin{tikzpicture}[scale=5]
      \draw[very thick,-,red] (.4,0) node[below,black] {$\U_0$}
      -- (0,.1) node[below,black] {$\Ub$};
      \draw[very thick,-,red] (0,.1) -- (0,.9);
      \draw[very thick,-,red] (0,.9) -- (-.4,1) node[above,black] {$\U_1$};
      \draw[thick,|-|,black] (.2,1) -- (.2,.9)
      node[midway,right] {$\dz_1$};
      \draw[thick,|-|,black] (-.2,0) -- (-.2,.1)
      node[midway,left] {$\dz_0$};
    \end{tikzpicture}
  \end{center}
  \caption{Schematic representation of $\U_{\dt}(\z)$ for small $\dt$.  The
    temperature at $\z=0,1$ is $\U_{0,1}$, $\Ub$ is the temperature in the
    bulk, and the boundary layers are of thickness $\dz_{0,1}$.}
  \label{fig:Udtschem}
\end{figure}

\subsection{The internal temperature \texorpdfstring{$\Ub$}{Ubar} for rapid flipping}
\label{sec:U}

The presence of boundary layers with a constant bulk temperature allows us to
make a simple argument to predict the form of $\U_{\dt}$ as $\dt\rightarrow0$,
that is, when the flips are short.  \Cref{fig:Udtschem} shows schematically
what the limiting profile looks like, with a bulk temperature~$\Ub$ and
boundary layers of thickness $\dz$.  Since~$\U_{\dt}$ must satisfy
\cref{eq:BCT0,eq:BCT1}, we have
\begin{equation}
  \frac{\Ub - \U_0}{\dz_0} \approx -\h_0\,(1 - \U_0),
  \qquad
  \frac{\U_1 - \Ub}{\dz_1} \approx -\h_1\,\U_1,
  \label{eq:UBC}
\end{equation}
where we approximated derivatives by the slope in the boundary layers in
\cref{fig:Udtschem}.  We want the limit $\dz_i \rightarrow0$ to exist, so we
demand
\begin{equation}
  \Ub - \U_0 = \Order{\dz_0}
  \qquad
  \U_1 - \Ub = \Order{\dz_1},
  \qquad
  \dz_i \rightarrow 0.
\end{equation}
Subtracting these, we have
\begin{equation}
  \Ub = \tfrac12(\U_0 + \U_1) + \Order{\dz_i}.
\end{equation}
We can also write this as
\begin{equation}
  \U_0 = \Ub + \dU,
  \qquad
  \U_1 = \Ub - \dU.
\end{equation}
Returning to \cref{eq:UBC}, we have
\begin{equation}
  -\frac{\dU}{\dz_0} \approx -\h_0\,(1 - \U_0),
  \qquad
  -\frac{\dU}{\dz_1} \approx -\h_1\,\U_1.
  \label{eq:UBC2}
\end{equation}
The fluxes~$\dU/\dz_i$ in \cref{eq:UBC2} must be equal, since otherwise the
fixed-point profile would have a net gain or loss of heat during an
interval~$\dt$.  This means that~$\dz_0=\dz_1$ (the boundary layers have the
same thickness), and we can equate the two right-hand sides in \cref{eq:UBC2}
to obtain
\begin{equation}
  \Ub \approx \frac{\h_0}{\h_0 + \h_1},
  \qquad
  \dt\rightarrow0.
  \label{eq:Ub}
\end{equation}
For the parameters in \cref{tab:physical}, this gives $\Ub \approx 0.9375$, in
good agreement with the small-$\dt$ case in \cref{fig:flipheatfix}.  For
symmetric cases with~$\h_0=\h_1$, we always have~$\Ub=1/2$, as is apparent in
\cref{fig:flipheatfix_sym}.  Note that to maximize $\Ub$ in \cref{eq:Ub} we
should make $\h_1$ as small as possible for a given $\h_0$, that is, make the
top boundary as insulating as possible.  For~$\h_1=0$ (perfect insulator) we
have $\Ub=1$, independent of~$\h_0$.

The simple argument presented in this section must be taken with a grain of
salt.  It is clear from \cref{fig:flipheatfix_sym} that for large~$\h_i$ the
boundary layer drop~$\dU$ is not small
as~$\dt\rightarrow0$, so the argument is shaky for large~$\h_i$.  Empirically,
the formula~\eqref{eq:Ub} seems to hold nonetheless.  In that limit see
\cref{sec:blayer} for the exact form of the boundary layer.

\subsection{Eigenfunction formulation}
\label{sec:flipeigen}

The projection of the flipped eigenfunctions onto the upright (unflipped)
eigenfunctions is
\begin{equation}
  \flipp_{mn} =
  \langle{\eigff}_m,\eigf_n\rangle
  =
  \int_0^1{\eigff}_m(\z)\,\eigf_n(\z)\dint\z.
  \label{eq:flipp}
\end{equation}
The matrix~$\flipp$ is symmetric, as can be seen by putting~$\z' = 1-\z$
in \cref{eq:flipp}:
\begin{align*}
  \flipp_{mn} &=
  \int_0^1\eigf_m(1-\z)\,\eigf_n(\z)\dint\z 
  =
  \int_0^1\eigf_m(\z')\,\eigf_n(1-\z')\dint\z'
  =
  \flipp_{nm}.
\end{align*}
The Fourier matrix representation of the flip-heat operator~$\cK_{\dt}$
from~\eqref{eq:flipheat} is
\begin{align}
  {(K_{\dt})}_{mn}
  =
  \langle\cK_{\dt}\,\eigf_m,\eigf_n\rangle
  =
  \ee^{-\eigv_m^2\dt}\flipp_{mn}.
  \label{eq:Kmn}
\end{align}
Its transpose is the heat-flip operator, where we heat first for a time~$\dt$
and then flip.

The Fourier representation of \cref{eq:recsoleqdt} with~$\T_0(\z)=0$ is
\begin{equation}
  \hT_m(\t_k)
  =
  \sum_{n=1}^\infty
  [(\eye - K_{\dt})^{-1}\,(\eye - K_{\dt}^k)]_{mn}
  \,(1 - \ee^{-\eigv_n^2\dt})\,\hTs_n\,,
\end{equation}
which for large~$k$ converges to the fixed-point profile
\begin{equation}
  (\hU_{\dt})_m
  =
  \sum_{n=1}^\infty
  [(\eye - K_{\dt})^{-1}]_{mn}
  \,(1 - \ee^{-\eigv_n^2\dt})\,\hTs_n.
  \label{eq:Uh}
\end{equation}
Though these expressions are computationally useful, they are difficult to
use for analytical progress, except in the symmetric case that we treat in
\cref{sec:sym}.

\subsection{The spectrum of \texorpdfstring{$\cK_{\dt}$}{Kdt}}
\label{sec:spectrumK}

As mentioned in \cref{sec:flipfixed}, the rate of convergence of temperature
to the fixed-point profile~$\U_{\dt}$ is determined by the spectrum of the
flip-heat operator~$\cK_{\dt} = \cH_{\dt}\cF$.
Write~$\psi_m(\z)$ and~$\sigma_m$ for the eigenfunctions and eigenvalues
of~$\cK_{\dt}$:
\begin{equation}
  \cK_{\dt}\psi_m(\z) = \sigma_m\psi_m(\z).
\end{equation}
(We suppress the~$\dt$ dependence of~$\psi_m$ and~$\sigma_m$ to keep the
notation light.)
Let~$\chi_m(\z) = \cF\psi_m(\z)$ be the `flipped' eigenfunction:
\begin{equation}
  \cK_{\dt}\psi_m(\z) = \cH_{\dt}\chi_m(\z) = \sigma_m\cF\chi_m(\z).
\end{equation}
Multiplying by~$\chi_n(\z)$ and integrating, we have
\begin{equation}
  \langle\chi_n,\cH_{\dt}\chi_m\rangle
  =
  \sigma_m\langle\chi_n,\cF\chi_m\rangle.
\end{equation}
Because both~$\cH_{\dt}$ and~$\cF$ are self-adjoint, we can interchange~$m$
and~$n$ to get
\begin{equation}
  (\sigma_m - \sigma_n)\langle\chi_n,\cF\chi_m\rangle = 0.
\end{equation}
Hence, the eigenfunctions~$\psi_m(\z)$ are $\cF$-orthogonal:
\begin{equation}
  \langle\psi_m,\cF\psi_n\rangle = 0,
  \qquad
  \sigma_m \ne \sigma_n.
\end{equation}
The self-adjointness of~$\cH_{\dt}$ and~$\cF$ and the positivity
of~$\cH_{\dt}$ also imply that the~$\sigma_m$ are real.  Note that, unlike the
eigenvalues of~$\cH_{\dt}$, here eigenvalues alternate sign:
$\sigma_m = (-1)^{m+1}\,\lvert\sigma_m\rvert$.

%
\begin{figure}
  \begin{center}
    \includegraphics[width=.6\textwidth]{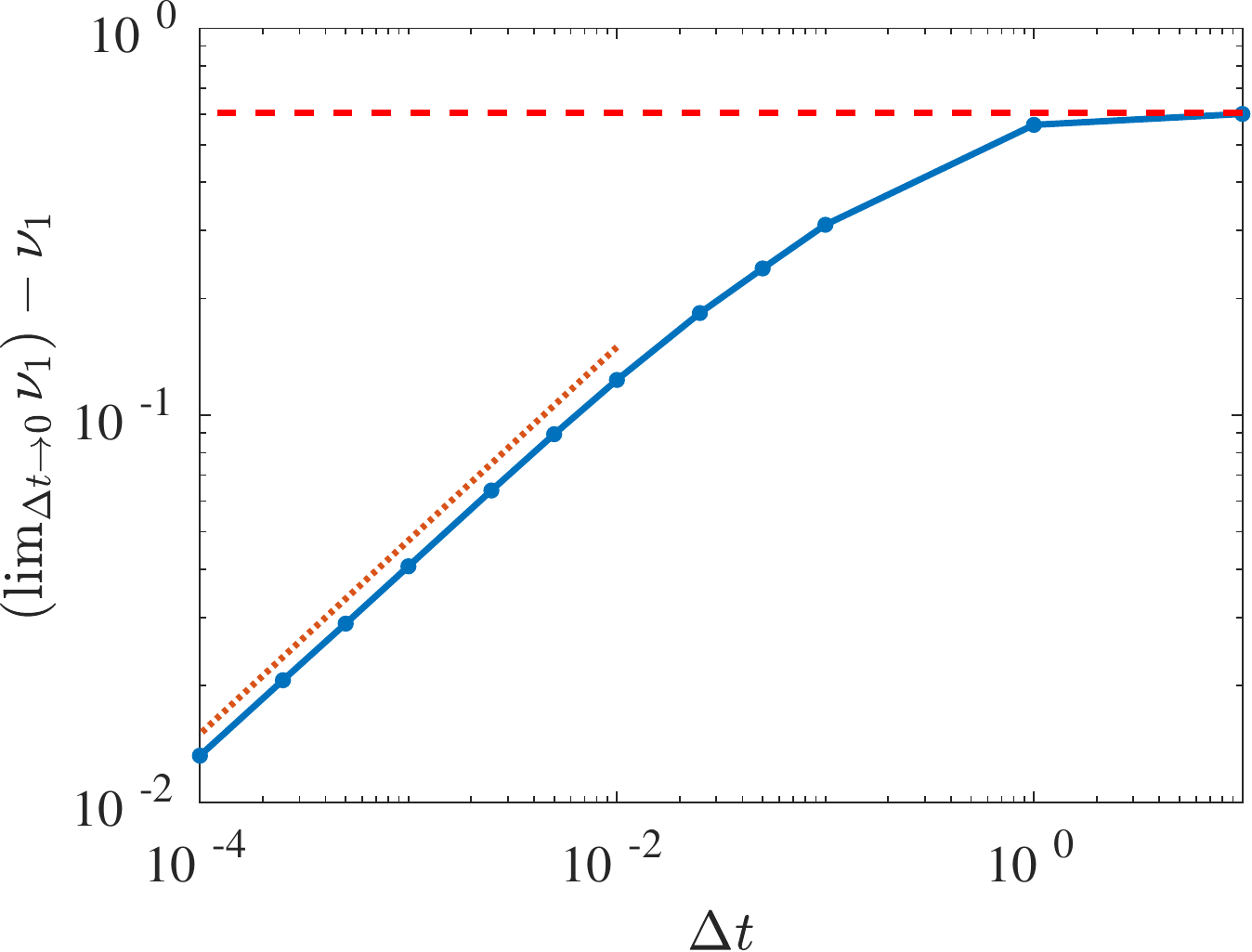}
  \end{center}
  \caption{Dependence on~$\dt$ of the leading decay rate~$\nu_1$
    [\cref{eq:nu}] for the flip-heat operator~$\cK_{\dt}$.  The limit
    of~$\nu_1$ as $\dt\rightarrow0$ is about~$2.685$, with convergence at a
    rate~$\sqrt{\dt}$ (dotted line).  The horizontal dashed line
    indicates~$\eigv_1$ for the heat operator.}
  \label{fig:flipheatop_spectrum_conv}
\end{figure}

Recall that the eigenvalues of the heat operator~$\cH_{\dt}$
are~$\ee^{-\eigv_m^2\dt}$.  In order to compare decay rates with and without
flipping, we define the decay rate~$\nu_m>0$ by
\begin{equation}
  \nu_m^2 = -\frac{1}{\dt}\log\,\lvert\sigma_m\rvert,
  \qquad
  m=1,2,3,\ldots.
  \label{eq:nu}
\end{equation}
The~$\nu_m$ can be compared to the~$\eigv_m$ to quantify how flipping
accelerates convergence to the fixed-point profile $\U_{\dt}$, versus
convergence to the steady solution~$\Ts(\z)$ in the absence of flipping.
Unlike~$\eigv_m$, $\nu_m$ depends on the flip interval~$\dt$.  For large~$\dt$
the $\nu_m$ converge to the~$\eigv_m$.  For small $\dt$, the limit is
nontrivial: in \cref{fig:flipheatop_spectrum_conv} we can see that~$\nu_1$
converges to approximately~$2.685$ as~$\dt\rightarrow0$, for the reference
parameter values in \cref{tab:physical}.  This is significantly faster than
the corresponding unflipped eigenvalue $\eigv_1 \approx 2.0803$ for the same
parameter values.  The acceleration ratio is thus
\begin{equation}
  \frac{\nu_1^2}{\eigv_1^2}
  \approx
  \frac{(2.685)^2}{(2.0803)^2} \approx 1.67,
  \qquad
  \dt\rightarrow0,
  \label{eq:nu1_ratio_dt0}
\end{equation}
that is, in the limit of very rapid flips, the approach to equilibrium is
about~$2/3$ faster.  This does not mean that the total cooking time is that
much faster, since it also depends on the profile~$\U_{\dt}$.  As we will see
in \cref{sec:sym}, for the symmetric case $\h_0=\h_1$ we have~$\nu_1=\eigv_1$,
and yet the cooking time is still much faster with flipping than without.  A
rigorous proof of the existence of the limit~\eqref{eq:nu1_ratio_dt0}, or an
estimate of its value, is an interesting open question.

\section{Symmetric case (\texorpdfstring{$\h_0=\h_1$}{h0=h1})}
\label{sec:sym}

Things simplify considerably for the up-down symmetric case~$\h_0=\h_1$.  The
system as a whole is still asymmetric in that it is heated at~$\z=0$ and
cooled at~$\z=1$, but the material properties are the same on the top and
bottom.  This could be achieved by placing on top of the food an unheated
metal plate with the same properties as the bottom heating plate.  For
instance, a top metal plate is used to make thin `smashed burgers.'  The
symmetric case is analogous to the Boussinesq limit for Rayleigh--B\'enard
convection, as opposed to the non-Boussinesq case which has non-symmetric
boundary conditions~\cite{Zhang1997}.

For this symmetric case, the flip and heat operators commute:
\begin{equation}
  \cF\cH_\t = \cH_\t\cF.
\end{equation}
The eigenfunctions~$\eigf_m(\z)$ of~$\cH_\t$ have the
parity property
\begin{equation}
  \cF\eigf_m(\z) = (-1)^{m+1}\eigf_m(\z).
\end{equation}
Hence, from \cref{eq:flipp},
\begin{equation}
  \flipp_{mn} = (-1)^{m+1}\,\delta_{mn}.
\end{equation}
The eigenvalues and eigenfunctions of~$\cK_\t = \cH_\t\cF$ are then
\begin{equation}
  \sigma_m = (-1)^{m+1}\,\ee^{-\eigv_m^2\t},
  \qquad
  \psi_m(\z) = (-1)^{m+1}\,\eigf_m(\z),
\end{equation}
and~$\nu_m=\eigv_m$ in \cref{eq:nu}.  Therefore, the \emph{rate} of convergence
to the fixed-point profile is not accelerated.  The improvement in cooking
time will arise because with flipping we expect that the final point to be
cooked is close to~$\z=1/2$, as opposed to~$\z=1$ without flipping
(\cref{sec:noflip}).

The steady-state profile~\eqref{eq:Ts} can be divided into a constant part
(even under $\cF$) and a part proportional to $\z -\tfrac12$ (odd under
$\cF$):
\begin{align}
  \Ts(\z) &= \tfrac12(\Ts(0) + \Ts(1)) - (\Ts(0) - \Ts(1))(\z - \tfrac12)
  \nonumber \\
  &= \tfrac12 - (1 + 2\h_0^{-1})^{-1}\frac{}{}(\z - \tfrac12)
  .
  \label{eq:Tsevenodd}
\end{align}

From \cref{eq:Uh}, we have for the fixed-point profile
\begin{align}
  \U_{\dt}(\z)
  &=
  \sum_{m=1}^\infty
  \frac{1 - \ee^{-\eigv_m^2\dt}}{1 + (-1)^m\,\ee^{-\eigv_m^2\dt}}
  \,\hTs_m\eigf_m(\z) \nonumber\\
  &=
  \sum_{\text{$m$ odd}}
  \hTs_m\eigf_m(\z)
  +
  \sum_{\text{$m$ even}}
  \tanh(\eigv_m^2\dt/2)
  \,\hTs_m\eigf_m(\z).
  \label{eq:Usym0}
\end{align}
The first sum is equal to the even (constant) part of the temperature
profile~\eqref{eq:Tsevenodd}, so~\eqref{eq:Usym0} simplifies to:
\begin{equation}
  \U_{\dt}(\z)
  =
  \tfrac12
  +
  \sum_{\text{$m$ even}}
  \tanh(\eigv_m^2\dt/2)
  \,\hTs_m\eigf_m(\z).
  \label{eq:Usym}
\end{equation}
We then have~$\U_{\dt}\bigl(\tfrac12\bigr) = \tfrac12$,
since~$\eigf_m\bigl(\tfrac12\bigr)=0$ for~$m$ even.  This is the type of
fixed-point profile plotted in \cref{fig:flipheatfix_sym}
for~$\h_0=\h_1=\infty$.  \keep{Show $\tanh$ part zero except in BLs?  Can use
  large-$m$ form of $\eigv_m \sim (m-1)\pi$.  The $\tanh$ acts as a high-pass
  filter.  Maybe show first it converges weakly to zero as $\dt\rightarrow0$.
  Do the $\h_0=\infty$ case, for which $\eigv_m = m\pi$.}  In
\cref{sec:blayer} we determine the boundary layer structure from
\cref{eq:Usym} in the limit of small~$\dt$.

Because~$\cF$ and~$\cH_{\dt}$ commute, the recurrence solution to the
flip-heat map \eqref{eq:recsoleqdt} after~$k$ flips is
\begin{equation}
  \T(\z,k\dt)
  =
  \U_{\dt}(\z)
  -\sum_m
  (-1)^{k(m+1)}\,
  \ee^{-\eigv_m^2 k\dt}\,(\widehat{U}_{\dt})_m\,\eigf_m(\z),
  \label{eq:Trecsolsym}
\end{equation}
\keep{Add a partial time interval at the end?  But the independence on $k$
  \emph{also} works when we end exactly at an interval.  Update: I don't quite
  remember what I meant by this.} where we take~$\T(\z,0)=0$ (food initially
uniformly cold).  Intuition suggests that the hardest point to cook is the
center~$\z=1/2$; assuming that this is so (we will discuss this assumption
below), the cooking time is obtained by solving
\begin{equation}
  \Tcook =
  \T\bigl(\tfrac12,\tcook\bigl)
  =
  \tfrac12 - \sum_{\text{$m$ odd}}
  \ee^{-\eigv_m^2 \tcook}\,\hTs_m\eigf_m\bigl(\tfrac12\bigr)
  \label{eq:tcooksym}
\end{equation}
for~$\tcook = k\dt$.  This cannot be solved analytically, but let
us see about finding~$\tcook$ by retaining only the slowest-decaying
mode~$m=1$:
\begin{equation}
  \tcook
  \approx
  \frac{1}{\eigv_{1}^2}
  \log
  \biggl(
  \frac
  {\hTs_{1}\eigf_{1}\bigl(\tfrac12\bigr)}{\tfrac12 - \Tcook}
  \biggr),
  \qquad
  \ee^{-(\eigv_3^2 - \eigv_1^2)\,\tcook} \ll 1.
  \label{eq:tcooksym1}
\end{equation}
Compare with \cref{eq:tcookthrough} for the cookthrough time.

Even though the cooking time~$\tcook$ is fairly short, the approximation
\eqref{eq:tcooksym1} is remarkably accurate because the error depends on the
ratio between~$\ee^{-\eigv_3^2\,\tcook}$ and~$\ee^{-\eigv_1^2\,\tcook}$:
eigenmodes that are odd under~$\cF$ do not contribute to the temperature at
the midpoint.  For~$\h_0=\h_1=\infty$, \cref{eq:tcooksym1} gives
\begin{equation}
  \tcook
  \approx
  \frac{1}{\pi^2}
  \log\biggl(
  \frac
  {2/\pi}{\tfrac12 - \Tcook}
  \biggr) \approx 0.097568,
  \label{eq:tcooksym_inf}
\end{equation}
whereas solving \cref{eq:tcooksym} numerically gives
$\tcook \approx 0.097584$.%
\matlabfootnote{See \MATLAB\ program \matlabcommand{tcooksym}.}
What is striking about \cref{eq:tcooksym1} is that it does not depend on the
number of flips, $k-1$, but only on the total time~$k\dt$.  This is clearly
due to the commutativity of~$\cH_{\dt}$ and~$\cF$.

\begin{figure}
  \begin{center}
    \subfigure[]{%
      \includegraphics[width=.468\textwidth]{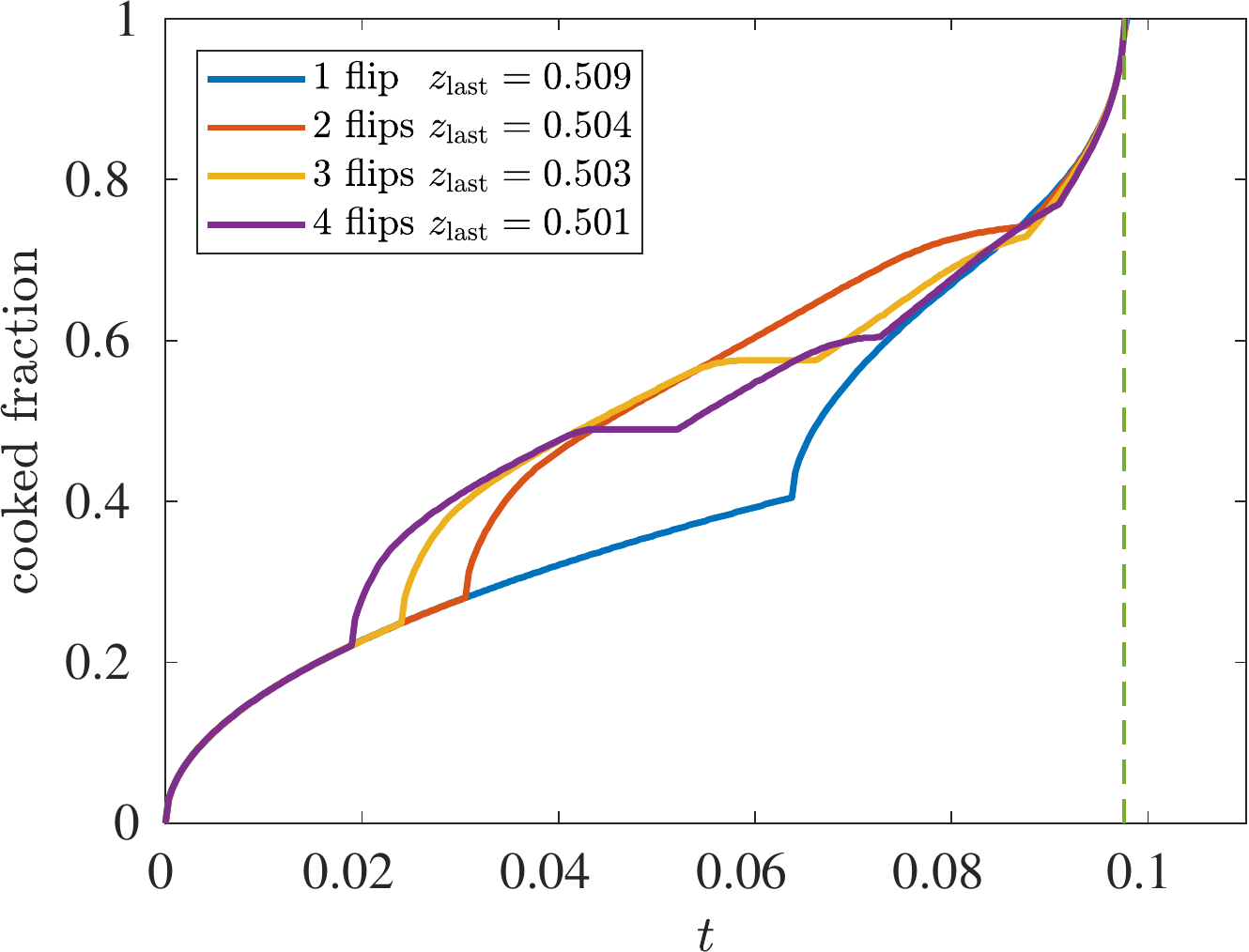}
      \label{fig:mincooktime_sym}
    }\hspace{.02\textwidth}
    \subfigure[]{%
      \includegraphics[width=.45\textwidth]{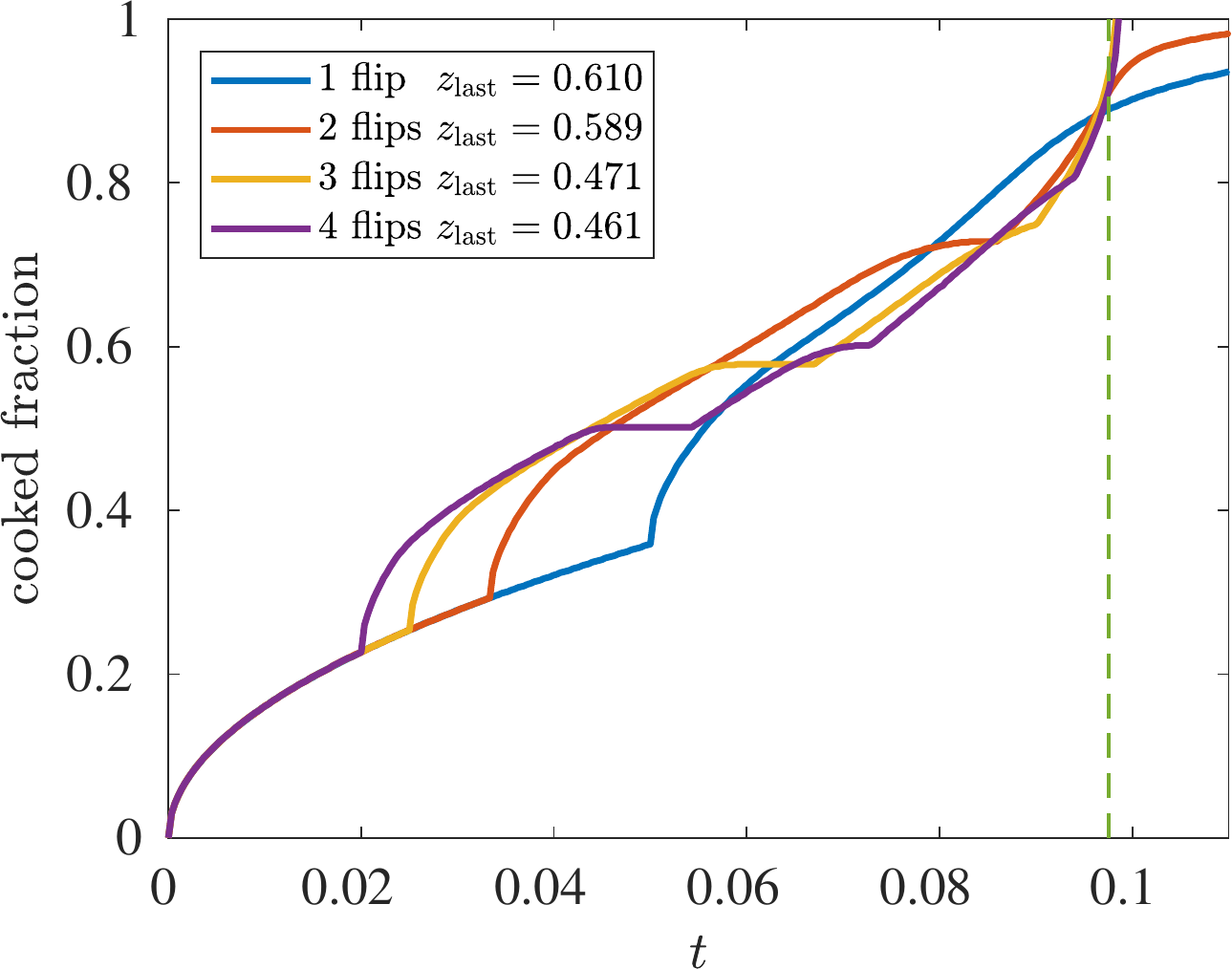}
      \label{fig:cooktime_sym}
    }
  \end{center}
  \caption{For symmetric perfectly-conducting boundary conditions
    $\h_0=\h_1=\infty$: (a)~Cooked fraction as a function of time for optimal
    solutions for~$1$ to~$4$ flips.  All the optimal times are essentially the
    same ($0.0976$), independent of the number of flips, and the final point
    to be cooked is always $\z \approx 0.5$.  (b) Cooked fraction for
    equal-time flips~$\dt = 0.1 / k$, where $k-1$ is the number of flips.  The
    vertical dashed line is the numerically-obtained optimal cooking time
    $0.0976$.}
  \label{fig:cooktime_sym_both}
\end{figure}

Can this be true?  It seems counter-intuitive that the cooking time is
completely independent of how we choose to flip, as long as we flip at least
once.  Yet numerical simulations support this, with an important caveat.  In
\cref{fig:mincooktime_sym} we show the `cooked fraction' (defined more
precisely in \cref{sec:opt}) as a function of time, for varying numbers of
flips.  These solutions all manage to cook the food in a time of~$0.0976$!
And yet in \cref{fig:cooktime_sym} we show a different set of flips, and two
cases take longer to cook.  The explanation is likely that
formula~\eqref{eq:tcooksym_inf} assumes that the last point to be cooked
is~$\z=1/2$, whereas for these two simulations this point is quite far
from~$\z=1/2$.

\keep{%
To ensure that the final point to be cooked is actually~$\z=1/2$, we add a
`tangency condition' that $\T_\z(\tfrac12,\tcook) = 0$; from
\cref{eq:Trecsolsym},
\begin{align}
  \T_\z\bigl(\tfrac12,\tcook\bigr)
  &=
  \U_{\dt}'\bigl(\tfrac12\bigr)
  -\sum_{\text{$m$ even}}
  (-1)^{k(m+1)}\,
  \ee^{-\eigv_m^2 \tcook}\,(\widehat{U}_{\dt})_m\,\eigf_m'\bigl(\tfrac12\bigr) \\
  &=
  \sum_{\text{$m$ even}}
  \l(1
  - (-1)^{k}\, \ee^{-\eigv_m^2 \tcook} \r)
  \tanh(\eigv_m^2\dt/2)\,\hTs_m\,\eigf_m'\bigl(\tfrac12\bigr)\\
\end{align}
}

\keep{This is just bizarre\dots Formula \cref{eq:tcooksym1} works perfectly
  \emph{as long as the final point to be cooked is $\z=1/2$}.  This seems to
  happen for $k \ge 4$, \ie, 3 flips or more.  In fact I suspect (and I think
  numerics confirm?) it will happen even without the equal flips, as long as
  the final cooked point is near $1/2$.  It's all due to the commutativity.
  The key is the ``tangency condition'' that the final point cooked also be
  tangent at~$1/2$?}

\section{Optimizing the flipping times}
\label{sec:opt}

\newcommand{\indf}[1]{\l\llbracket #1\r\rrbracket}
\mathnotation{\cookfrac}{\text{cooked fraction}}

In the previous sections we used the Eulerian description, where the food is
flipped and the coordinate~$\z$ refers to a fixed point in space.  The
Lagrangian or `co-flipping' description is in terms of a material
point~$\z_0 \in [0,1]$, which corresponds to a fixed point inside the food.
We can convert between the two pictures easily using the flip
operator~\eqref{eq:flip}:
\begin{equation}
  \z = (\cF)^{\#\text{flips}(\t)}\z_0\,,
\end{equation}
where~$\#\text{flips}(\t)$ gives the number of flips until time~$\t$.  Thus
the two reference frames coincide until the first flip, after which~$\z_0 =
1-\z$ until the next flip, etc.

In this section we will seek to optimize the time intervals where flips occur,
in order to find the optimal cooking time.  For this the Lagrangian
description is the appropriate one, since we must keep track of the history of
each material point in the food.  With a small abuse of notation we will
denote the temperature at material point~$\z_0$ at time~$\t$
by~$\T_0(\z_0,\t)$.  (Recall that~$\T_0(\z)$ was the initial temperature
profile.)

Note that in \cref{sec:sym} we estimated a cooking time by finding when the
center point \hbox{$\z=1/2$} was cooked.  Since this point is the same in the
Eulerian or Lagrangian frames, we did not have to change frames.

\subsection{Cooking time for variable intervals}
\label{sec:variable_dt}

Recall that we declare the food to be cooked if every point~$\z \in [0,1]$ has
at some time in its history achieved the temperature~$\Tcook$.  We define the
cooked fraction at time~$\t$ as
\begin{equation}
  \cookfrac(\t)
  =
  \int_0^1
  \indf{\max_{\s \in [0,\t]} \T_0(\z_0,\s) \ge \Tcook}
  \dint\z_0
  \label{eq:cookfrac}
\end{equation}
where~$\indf{A}$ is the indicator function of~$A$: it is one if~$A$ is true,
and zero otherwise.  The cooked fraction is a nondecreasing function of time,
since points cannot `uncook.'  The cooking time is then
\begin{equation}
  \tcook
  =
  \inf_{\t \ge 0}
  \l\{\cookfrac(\t) = 1\r\}.
  \label{eq:tcook}
\end{equation}
The cooking time could be infinite, as we saw in \cref{sec:noflip} when
cooking without flipping.

We specify a cooking protocol by fixing the number of cooking
intervals~$k \ge 1$ as well as the flipping times
\begin{equation}
  \dt_1,\dt_2,\ldots, \dt_{k-1},
  \qquad
  \dt_j > 0.
\end{equation}
Note that we do not allow any zero~$\dt_j$, as this would effectively reduce
to a cooking protocol with fewer flips.  We define the length of the final
cooking interval as
\begin{equation}
  \dt_k = \tcook - \sum_{j=1}^{k-1} \dt_j > 0.
\end{equation}

\begin{figure}
  \begin{center}
    \subfigure[]{%
      \includegraphics[height=.27\textheight]{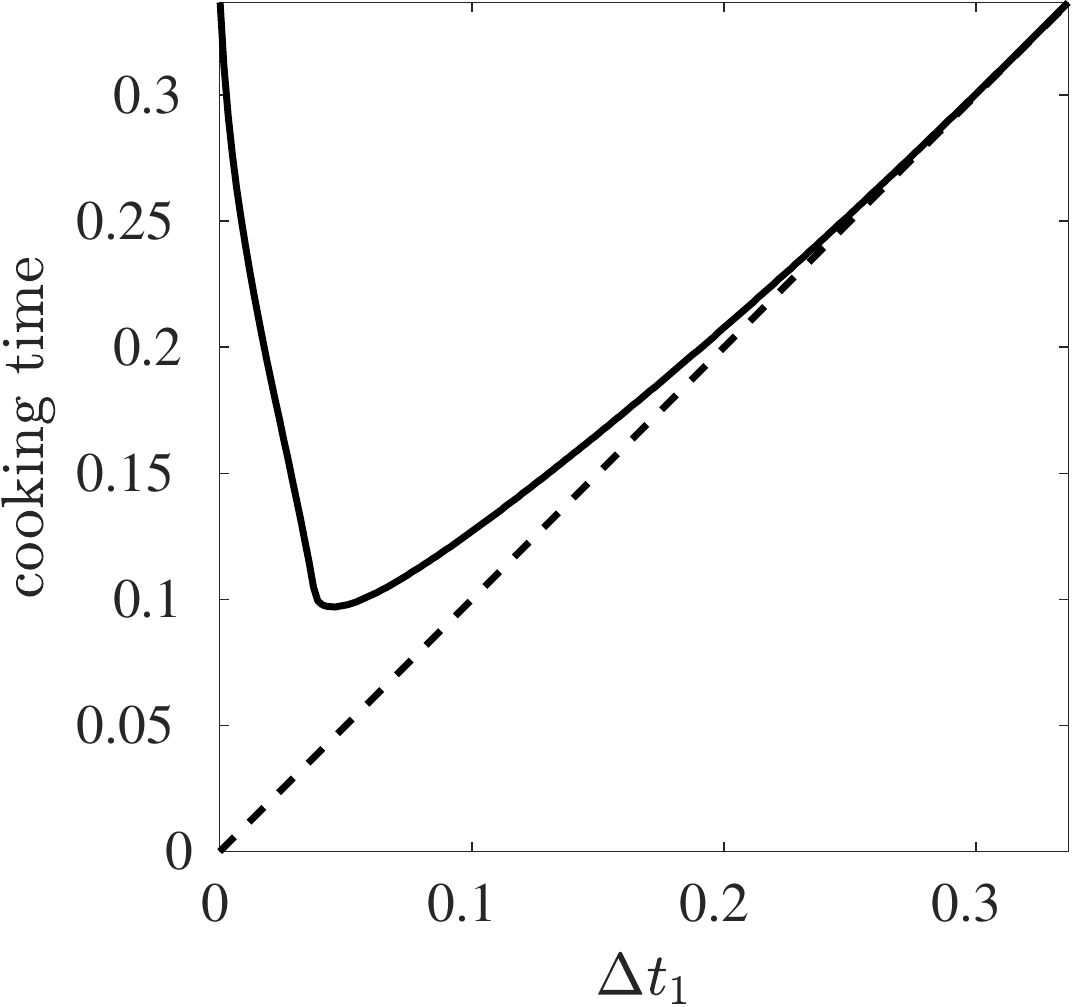}
      \label{fig:cooktimes_1flip}
    }\hspace{.02\textwidth}
    \subfigure[]{%
      \includegraphics[height=.29\textheight]{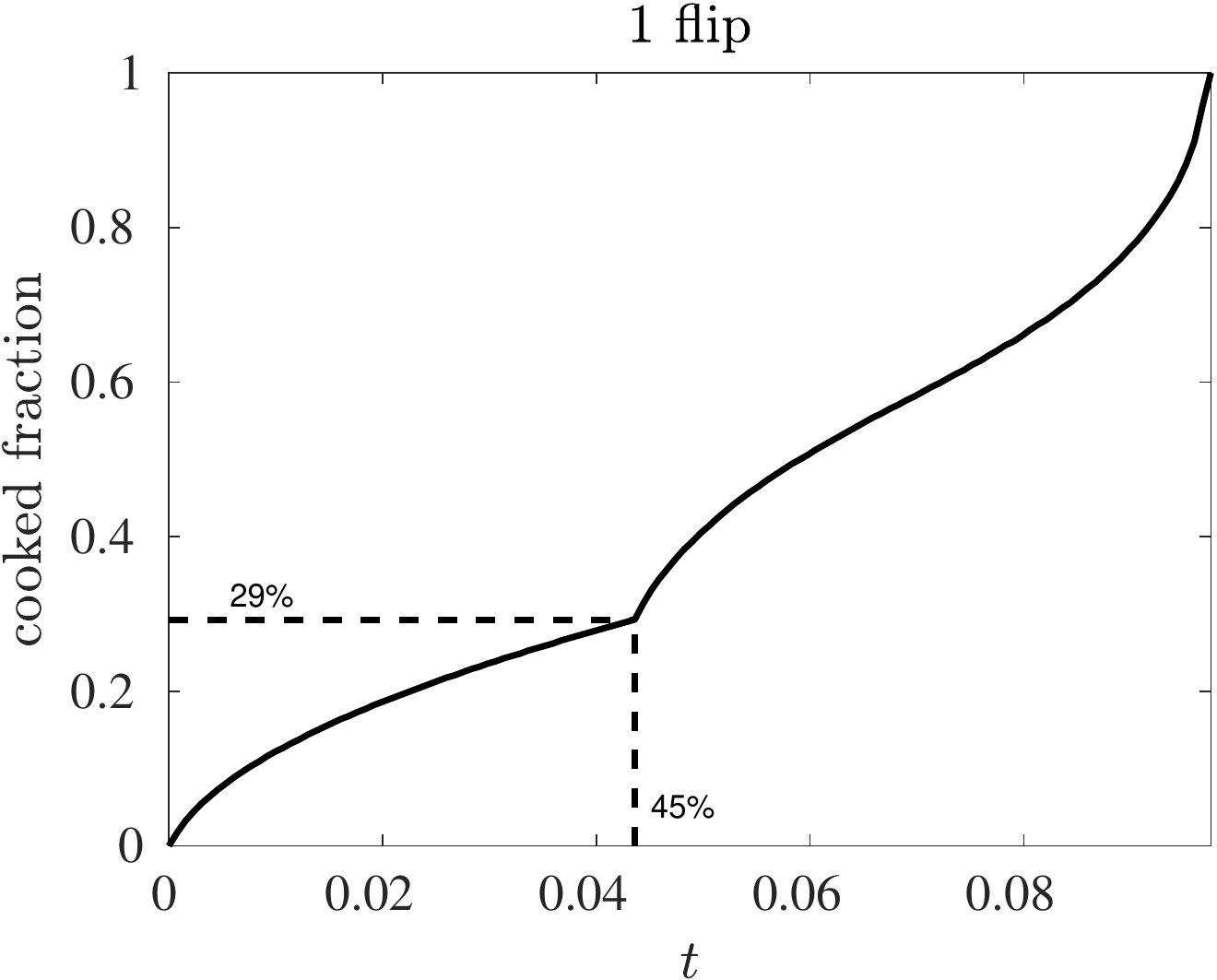}
      \label{fig:mincooktime_1flip}
    }
  \end{center}
  \caption{Cooking with one flip.  (a) The total cooking time~$\tcook$ as a
    function of the flipping time~$\dt_1$, showing a single minimum.  The
    dashed line is the diagonal $\tcook=\dt_1$; the solid curve crosses the
    dashed line at $\dt_1 = \tct \approx 0.340$ (\cref{sec:noflip}).  (b)
    Cooked fraction as a function of time for the optimal solution (the
    minimum in (a)).  The phase after the flip is longer, and there is rapid
    cooking at the very end.  The percentages indicate that the food is $29\%$
    cooked at the optimal flipping time $\dt_1 \approx 0.04359$, which occurs
    at $45\%$ of $\tcook$.}
  \label{fig:cooktimes_1flip_both}
\end{figure}

\Cref{fig:cooktimes_1flip} shows the cooking time~$\tcook$ as a function
of~$\dt_1$, for $k=2$ cooking intervals (1 flip).%
\matlabfootnote{See \MATLAB\ program \matlabcommand{cooktime}.}
The dashed line is the diagonal $\tcook=\dt_1$; since~$\tcook = \dt_1 + \dt_2$
with~$\dt_2>0$, the solid cooking time curve must lie above this diagonal.
The curve for~$\tcook$ intersects the diagonal as~$\dt_1 \rightarrow \tct$,
since for these parameters the cookthrough time is not infinite; otherwise
$\tcook$ asymptotes to the diagonal.  The cooking time~$\tcook$ also
asymptotes to~$\tct$ as~$\dt_1 \rightarrow 0$.  \Cref{fig:cooktimes_1flip}
suggests that there is a unique minimum cooking time for one flip,
$\tcook \approx 0.09751$.

The shape of~$\tcook$ near the minimum in \cref{fig:cooktimes_1flip} has an
important consequence: notice that~$\tcook$ is quite a bit steeper to the
left of the minimum than to the right.  This means that it is preferable to
err to the \emph{right} of the minimum (\ie, waiting a bit longer to flip),
since this does not increase the cooking time substantially.  Making~$\dt_1$
too small, however, will necessitate a longer final phase of cooking.

\begin{figure}[H]
  \begin{center}
    \subfigure[]{%
      \includegraphics[height=.27\textheight]{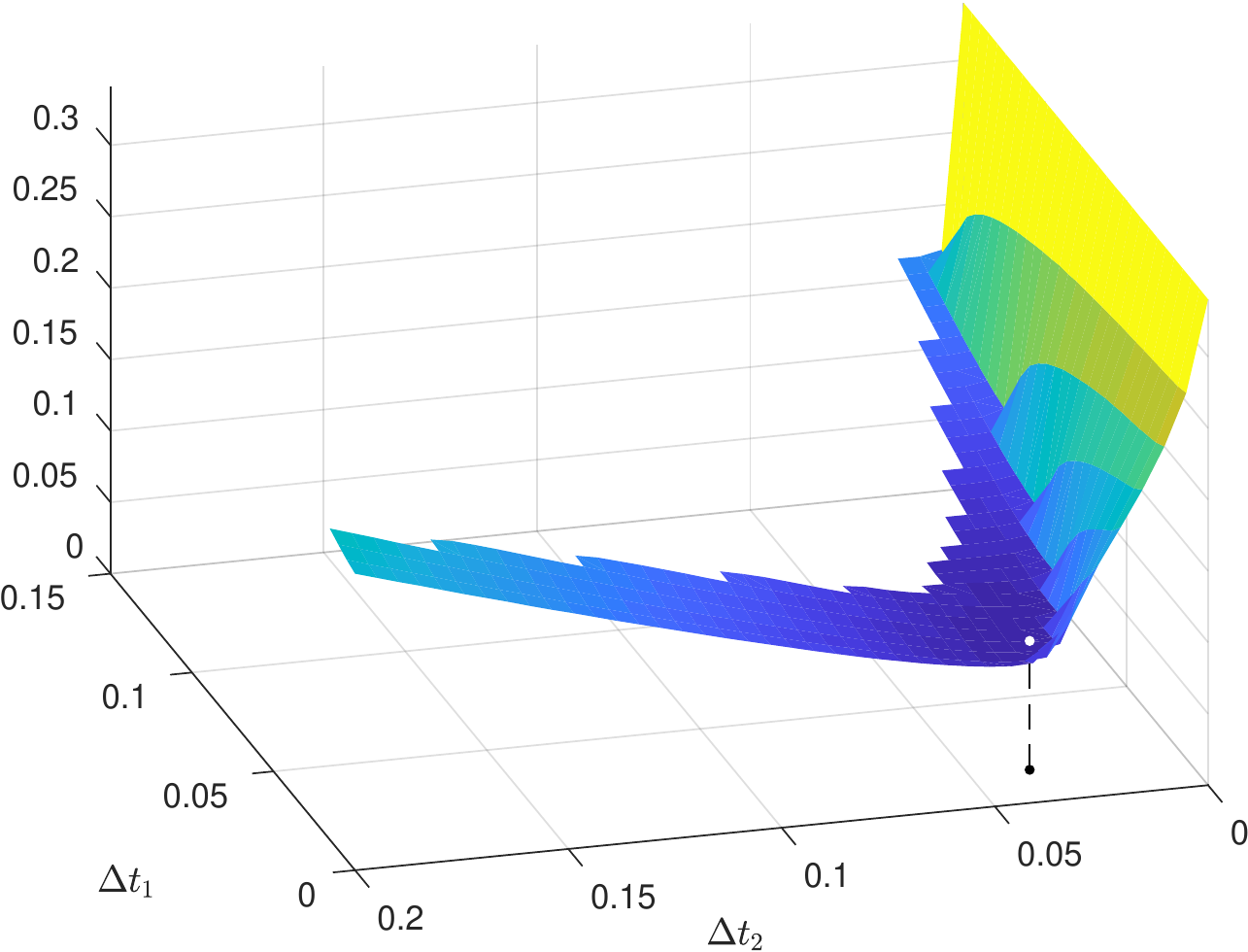}
      \label{fig:cooktimes_2flips}
    }\hspace{.02\textwidth}
    \subfigure[]{%
      \includegraphics[height=.29\textheight]{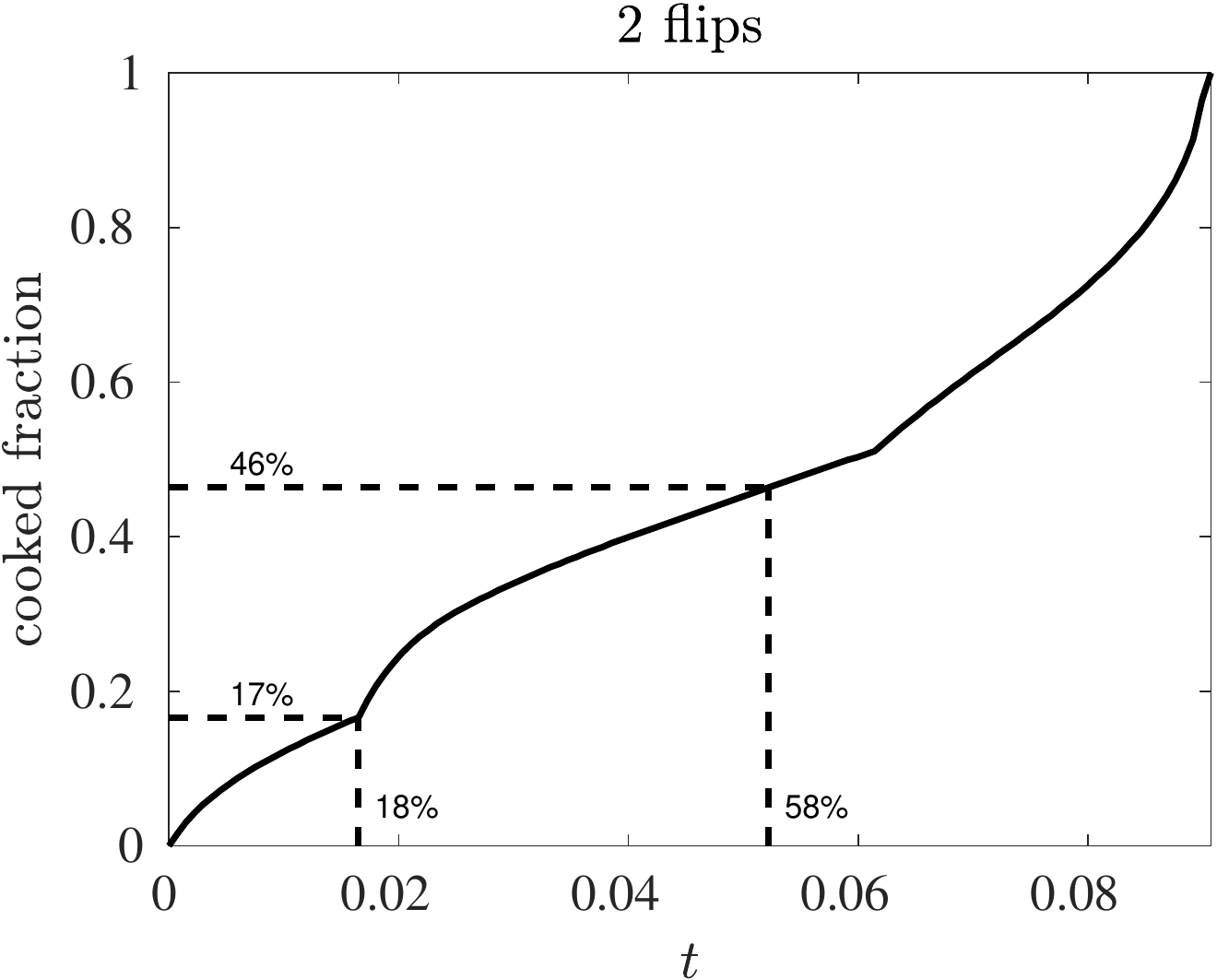}
      \label{fig:mincooktime_2flips}
    }
  \end{center}
  \caption{Cooking with two flips.  (a) The total cooking time as a function
    of the length the two flipping intervals~$\dt_1$ and~$\dt_2$, again
    showing a well-defined unique minimum.  In the region without data the
    food is cooked before a second flip, so the region does not count as two
    flips.  (b) Cooked fraction as a function of time for the optimal 2-flip
    solution.  There is an $8\%$ improvement in cooking time over a single
    flip.  The time~$\dt_1$ is quite short, making up only~$18\%$ of the total
    cooking time.  Note that the times with discontinuous slopes are not
    necessarily the flipping times.}
  \label{fig:cooktimes_2flips_both}
\end{figure}

\Cref{fig:cooktimes_2flips} shows the cooking time~$\tcook$ if we flip the
food twice, with three time intervals~$[0,\dt_1)$, $[\dt_1,\dt_1+\dt_2)$,
and~$[\dt_1+\dt_2,\tcook]$ .  In the same manner as we saw for one flip, the
figure suggests a unique global minimum.  In the regions without data the food
is cooked before the second flip occurs, so we ignore those values (\ie, we
plot nothing).  As for the single-flip case, the steepness of the minimum
suggests that we should cook slightly longer than the optimal values, since an
error on the left has larger consequences.

\subsection{Numerical optimization}
\label{sec:numopt}

Because of the nonlocal-in-time nature of the objective
function~\eqref{eq:tcook}, optimizing for the flipping times is difficult and
necessitates a numerical approach.  Furthermore, this optimization problem is
likely nonconvex and we have no guarantees that we have found the global
minimum.  Despite this, the swift convergence and robustness of the code
suggests that the minima we find are most likely global.

We use \MATLAB's nonlinear optimization function \matlabcommand{fminsearch} to
look for combinations of cooking intervals that lead to the fastest total
cooking time.%
\matlabfootnote{See \MATLAB\ program \matlabcommand{mincooktime}.}
The optimal solution for~$1$ flip ($k=2$ two cooking intervals) is plotted in
\cref{fig:mincooktime_1flip}.  We show the cooked fraction as a function of
time, so we can see how much each cooking interval contributes to the total
cooking.  The first interval is somewhat shorter than the second, and it leads
to far less cooking: the first interval is~$45\%$ of the total time, but leads
only to~$29\%$ of the cooking.  There is also a sudden spurt of cooking at the
end of the second interval.  This feature is present in optimal solutions with
more flips as well, and suggests that an optimal solution seeks to build up a
large temperature gradient that then diffuses and cooks rapidly.

The optimal solution for~$2$ flips ($k=3$ two cooking intervals) is plotted in
\cref{fig:mincooktime_2flips}.  Again the first interval is relatively short,
and does not lead to much cooking.  The final interval is long ($42\%$ of
total time), and gives a whopping~$54\%$ of the cooking.  Observe that the
curve has a point of discontinuous slope that doesn't coincide with a flip.
This can be traced to the $\max$ function in our definition of the cooked
fraction~\eqref{eq:cookfrac}: when we flip the food, points that were on the
top surface have cooled down, but have not uncooked.  Some time is needed for
the heat to propagate into the interior from the hot plate to resume cooking.

\begin{figure}
  \begin{center}
    \subfigure[]{%
      \includegraphics[height=.27\textheight]{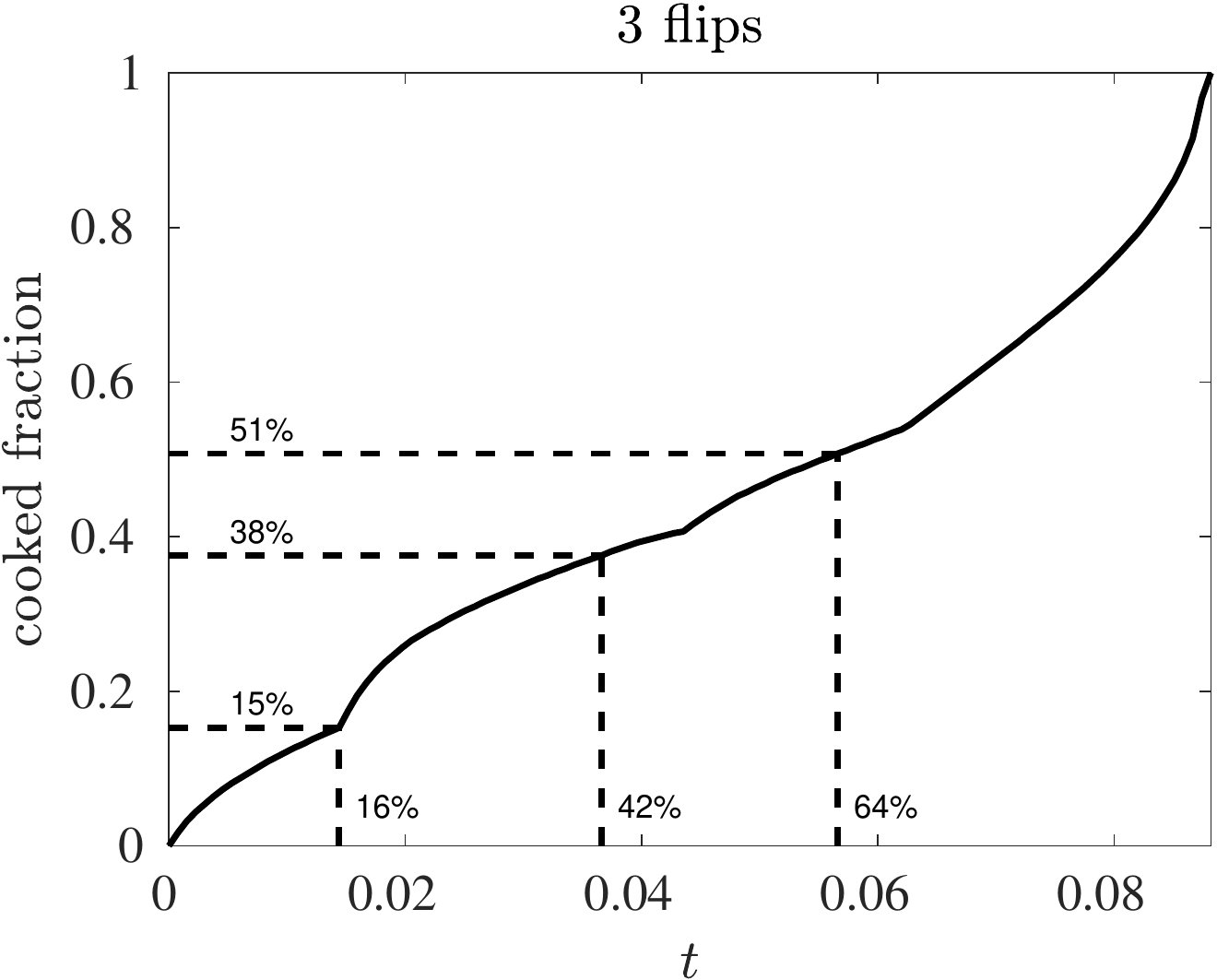}
      \label{fig:mincooktime_3flips}
    }\hspace{.02\textwidth}
    \subfigure[]{%
      \includegraphics[height=.27\textheight]{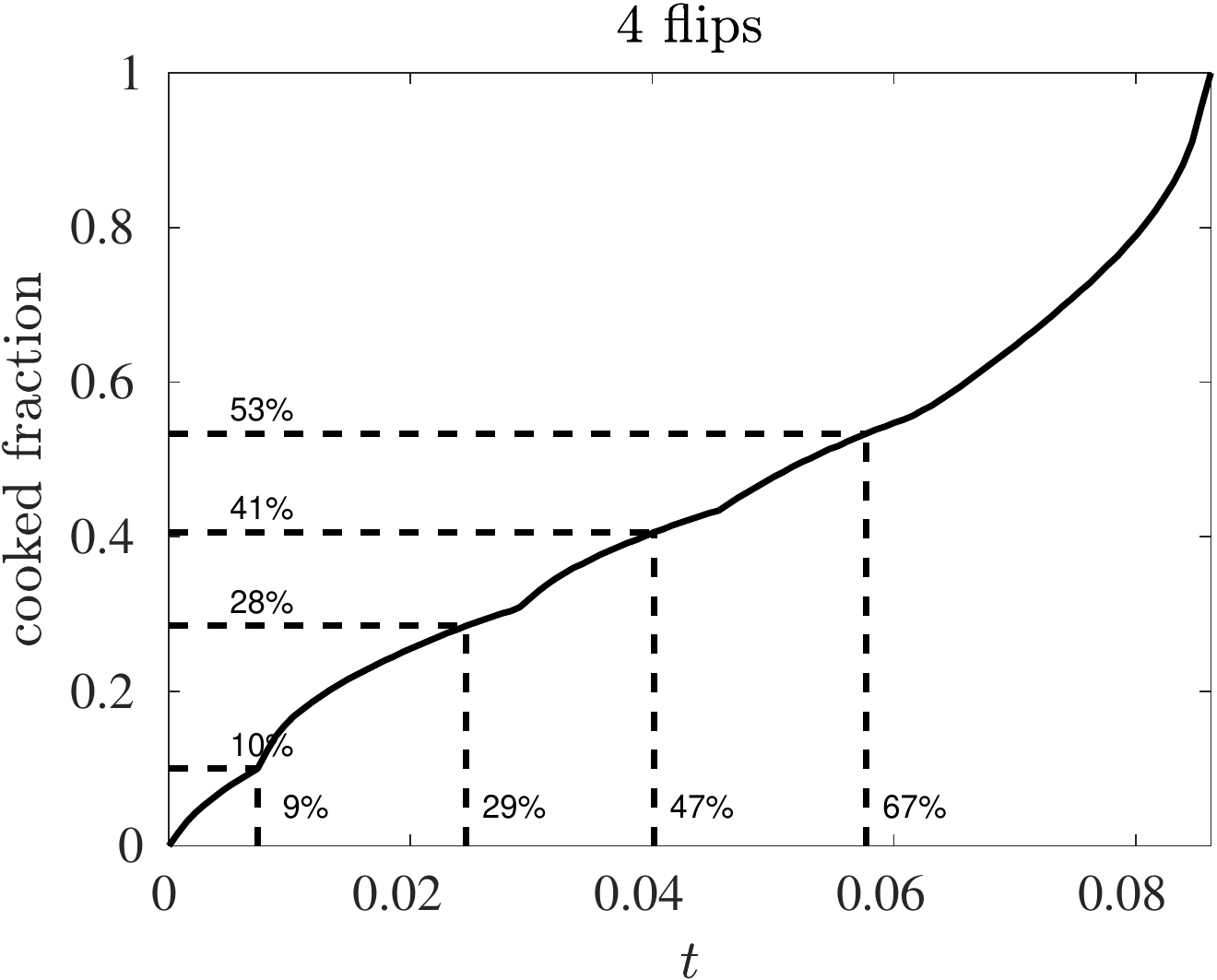}
      \label{fig:mincooktime_4flips}
    }

    \subfigure[]{%
      \includegraphics[height=.27\textheight]{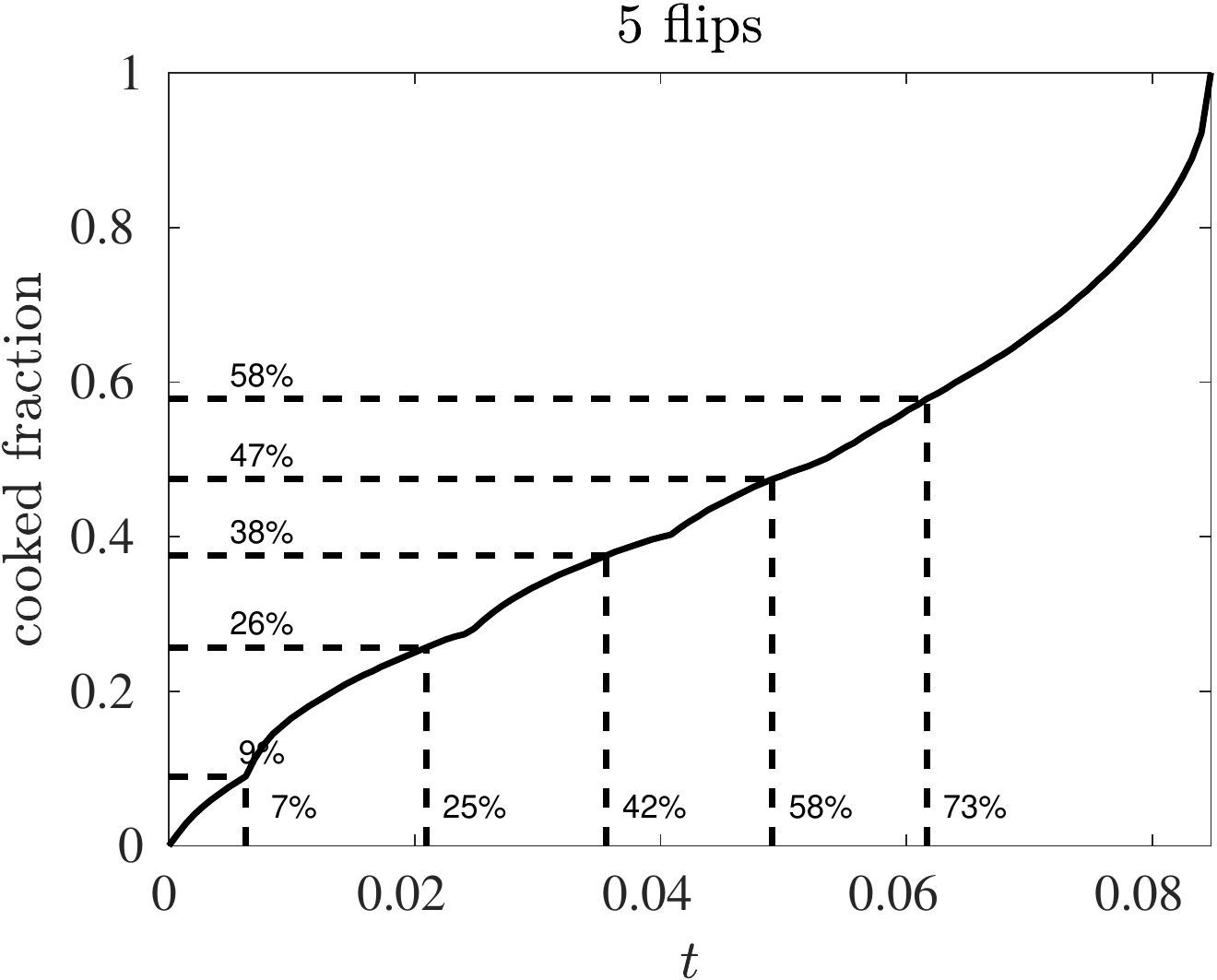}
      \label{fig:mincooktime_5flips}
    }\hspace{.02\textwidth}
    \subfigure[]{%
      \includegraphics[height=.27\textheight]{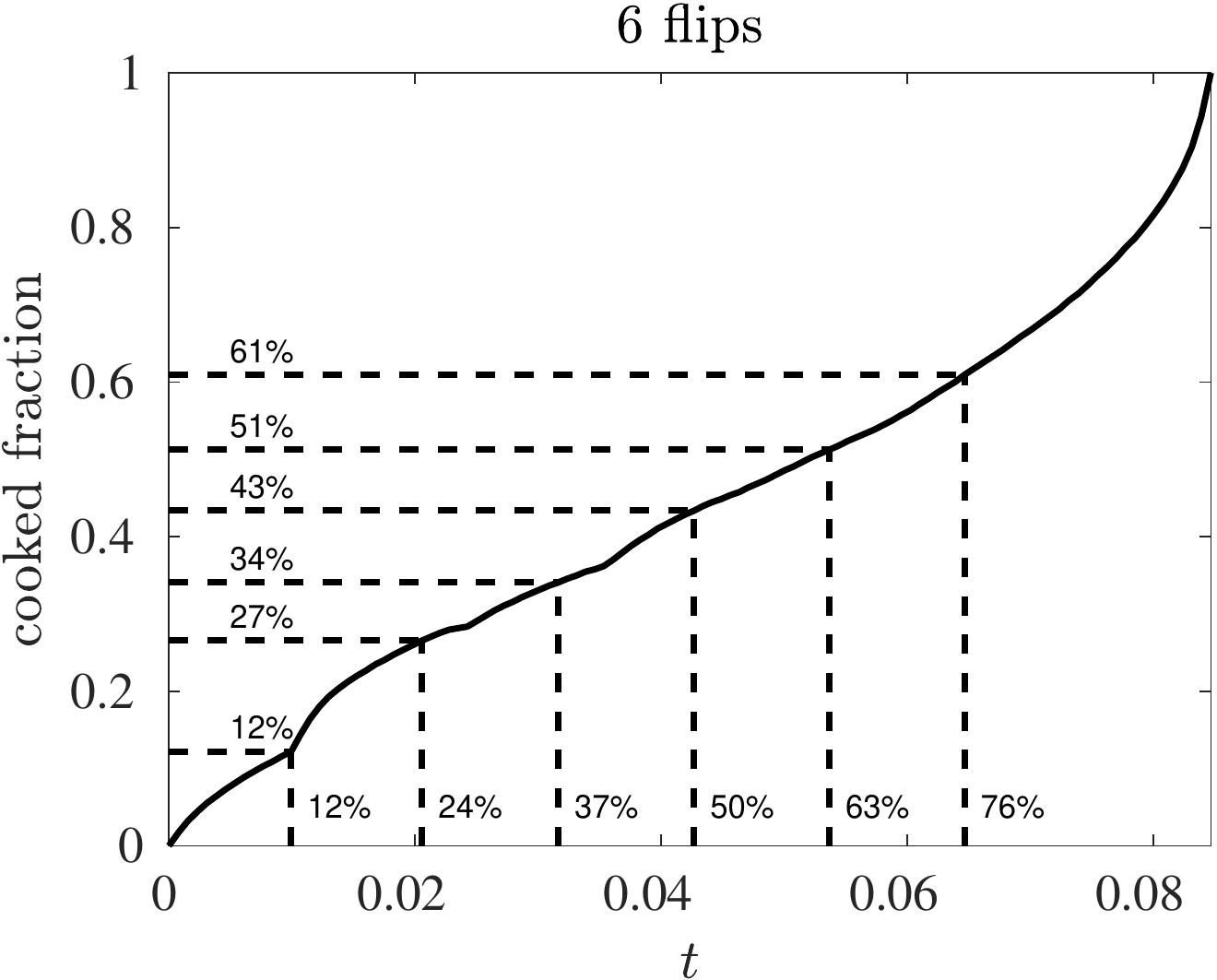}
      \label{fig:mincooktime_6flips}
    }

  \end{center}
  \caption{(a)--(d) Optimal flipping times for $3$--$6$ flips.  In each case
    the initial time~$\dt_1$ is fairly short, and the last interval is much
    longer.  The improvement with more flips is marginal (see
    \cref{fig:mincooktimes}).}
  \label{fig:mincooktime_Nflips}
\end{figure}

\begin{figure}
  \begin{center}
    \subfigure[]{%
      \includegraphics[height=.27\textheight]{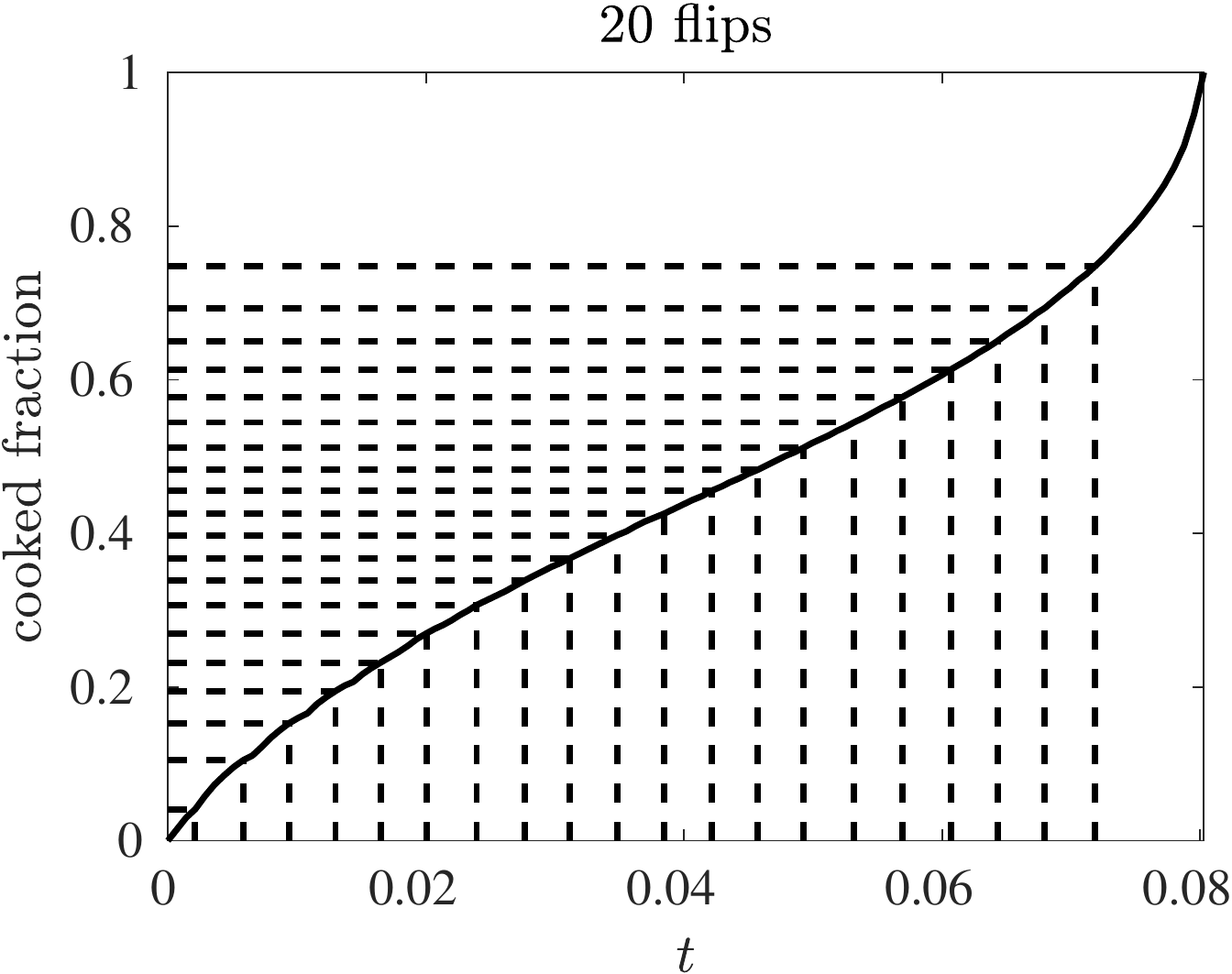}
      \label{fig:mincooktime_20flips}
    }\hspace{.02\textwidth}
    \subfigure[]{%
      \includegraphics[height=.265\textheight]{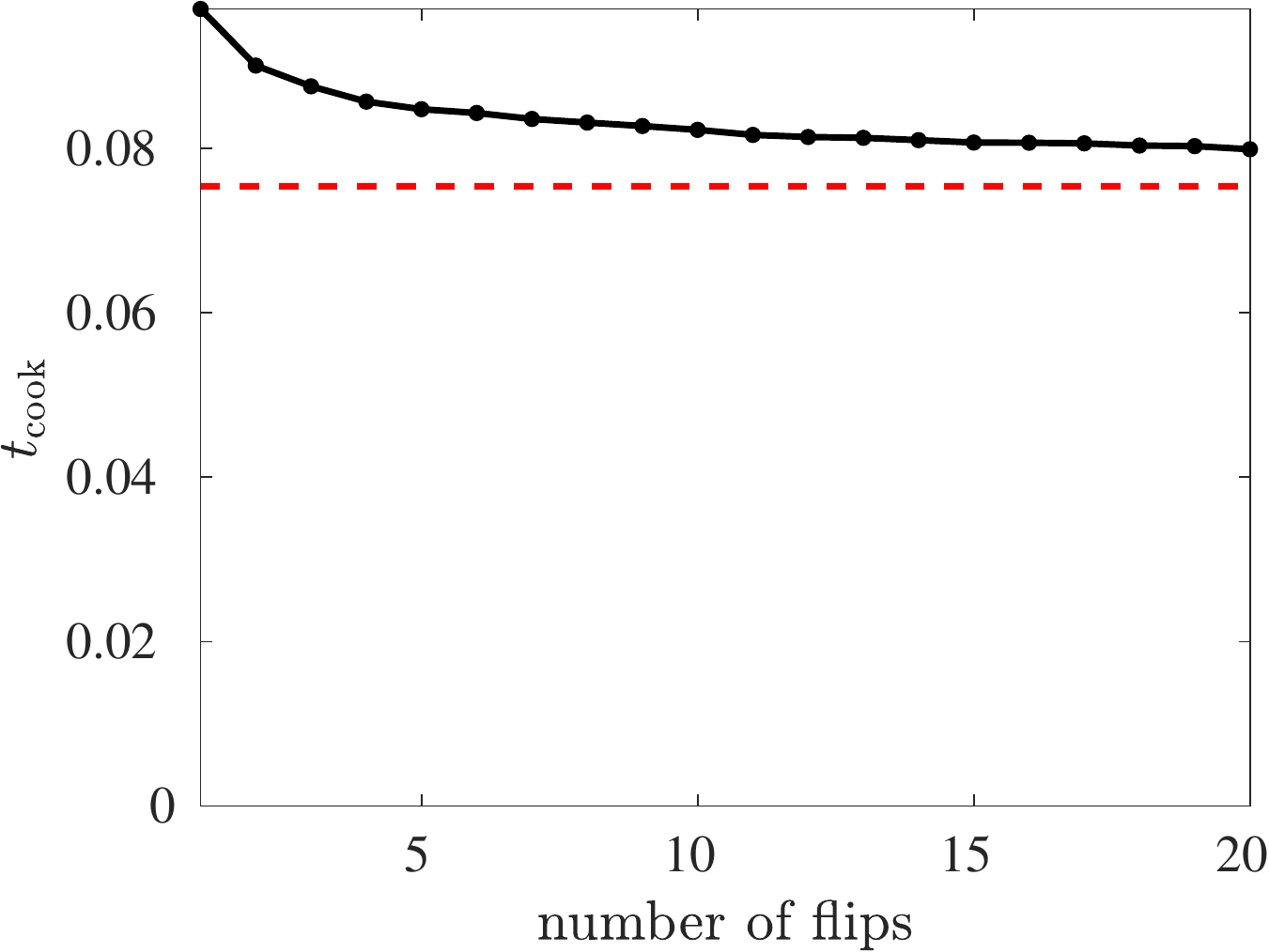}
      \label{fig:mincooktimes}
    }
  \end{center}
  \caption{(a) The optimal times for~$20$ flips: the intervals now have very
    comparable lengths, except the final one which is about twice as long.
    (b) For a large number of flips, the optimal cooking time~$\tcook$
    converges to $\approx 0.0754$ ($63\,\second$), as compared to $0.0970$
    ($80.5\,\second$) for a single flip.  This is a theoretical maximum
    decrease of $29\%$ (close to the Food Lab
    prediction~\cite{FoodLab_flipping}, see \cref{sec:intro}).}
  \label{fig:mincooktime_20flips_and_mincooktimes}
\end{figure}

In \cref{fig:mincooktime_Nflips} we increase the number of flips gradually.
We observe that the cooking intervals become more evenly spaced, except for
the final one.  We push this to the limit in \cref{fig:mincooktime_20flips}:
there we show the optimal solution for~$20$ flips, which no chef should
attempt.  The intervals are very similar in length, except again for the final
one.  The improvement in optimal cooking time as a function of the number of
flips, shown in \cref{fig:mincooktimes}, is fairly marginal after a few flips.
The optimal cooking time time appears to asymptote to a nonzero value.

\begin{figure}
  \begin{center}
    \includegraphics[height=.7\textwidth]{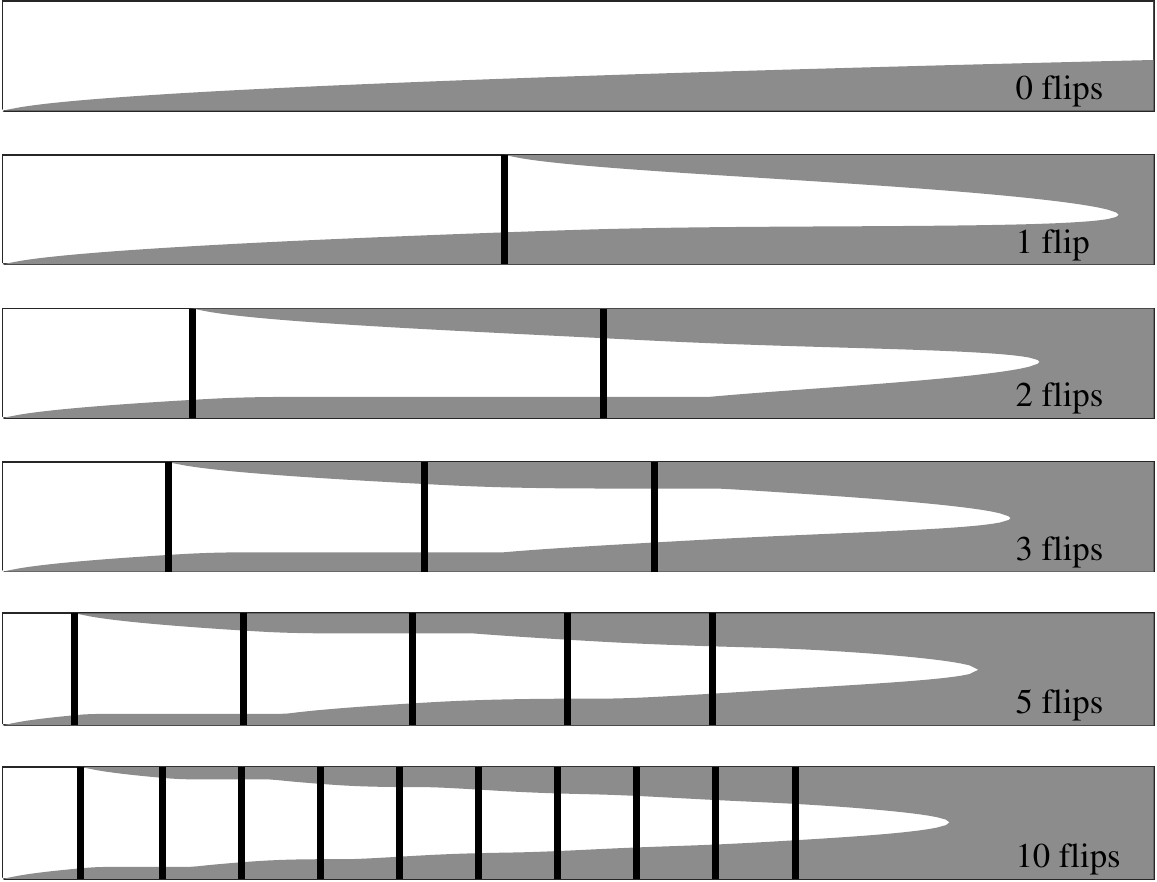}
  \end{center}
  \caption{Comparison of optimal solutions for different number of flips.  The
    vertical axis is the Lagrangian coordinate~$\z_0 \in [0,1]$, and the
    cooked region is shown in gray as a function of time.  The vertical lines
    indicate times of flipping.  We can clearly see that the surface initially
    at the top does not begin cooking until the first flip.  The time interval
    depicted (horizontal axis) is~$\t \in [0,0.1]$. (Parameters as in
    \cref{tab:physical}.)}
  \label{fig:cooking_history}
\end{figure}

To help visualize the cooking process, we show in \cref{fig:cooking_history}
the cooked region in gray as a function of time for several optimal solutions.
The top frame is the cooked region without flipping, which will cook at
time~$\tct \approx 0.340$ (\cref{sec:noflip}).  The vertical is the Lagrangian
coordinate~$\z_0 \in [0,1]$.  In that figure we are ``co-flipping'' with the
food, \ie, we are following material points.

\section{Discussion}
\label{sec:discussion}

In this paper we examined a simple model of cooking and flipping.  We solved a
one-dimensional heat equation with general boundary conditions in the form of
Newton's law of cooling.  We expressed the solution as a generalized Fourier
series in Sturm--Liouville eigenfunctions.  We saw immediately that the food
might never fully reach its cooked temperature, unless it is flipped.

Flipping the food (\ie, turning it over on the hot surface) can be regarded as
applying a map~$\z \mapsto 1-\z$ to the temperature field.  Composing such an
instantaneous flip with an interval~$\dt$ of heating gives the flip-heat
operator.  The spectral properties of this operator are themselves of some
interest, but have only been scratched here.  There does appear to be a
residual effect of flipping as the intervals become shorter: in that limit the
leading eigenvalue~$\nu_1^2$ of the flip-heat operator is about $67\%$ larger
than the purely-thermal value~$\mu_1^2$, for our reference parameters.  This
acceleration of convergence is similar in spirit to turbulent mixing in
thermal convection.

Of more importance than the spectrum for the purpose of cooking is the fixed
point of the flip-heat operator, which represents the equilibrium temperature
after many flips.  It is the increase in the interior temperature for this
fixed point that is mostly responsible for the improved cooking (see
\cref{eq:Ub} for small~$\dt$).  A related idea arises in the mixing of fluid:
in the presence of sources and sinks, it is often more practical to modify the
target temperature distribution rather than the rate of approach to that
target~\cite{Thiffeault2012,Thiffeault2008}.

The symmetric case, where the thermal properties at the top and bottom are the
same, is most readily solvable.  It exhibits some peculiarities, such as the
independence on the duration of flipping intervals, as long as we flip at
least once, and the final point to be cooked is the midpoint.  This is only
determined \textit{a posteriori}, so it is an open question if a more
practical criterion can be found.  Note that the symmetric versus
non-symmetric cases are the analogues of the Boussinesq versus non-Boussinesq
cases in Rayleigh--B\'enard convection~\cite{Zhang1997}.

Finally, we carried out numerical optimization of the cooking intervals.  We
fixed the number of flips, and found the flipping times that lead to the
shortest total cooking time.  The criterion for cooking is that all the points
in the food exceed a certain temperature at some point in their history.
Surprisingly, the numerical optimization appears to converge to a unique
global minimum, though it is not obvious why this problem should behave like a
convex one.  For several flips, the optimal intervals of cooking are roughly
of the same length, except for the final interval, which is typically longer
and where much of the cooking actually occurs.  The shape of the cooking time
function for two and three flips suggests that it is better to err on the side
of cooking a bit longer for each flip, since a shorter interval leads to a
longer cooking time (that is, for two flips the derivative is much steeper to
the left of the minimum than to the right; see \cref{fig:cooktimes_1flip}).

We emphasize that our results are mostly qualitative.  In particular, they are
too short by roughly a factor of two from typical cooking times.  But
the~$29\%$ decrease in total cooking time is relatively close to the Food Lab
prediction~\cite{FoodLab_flipping} mentioned in our introduction.  A realistic
model of cooking should include many subtle effects, such as the change in
moisture content and the melting of
fat~\cite{Dagerskog1977,Mathijssen2022_preprint,Ou2007}.  Of course, any
partially analytic treatment as given here quickly becomes impossible as more
effects are included.


\section*{Acknowledgments}

This project was originally inspired by Persi Diaconis and Susan Holmes.  The
author thanks Charles Doering for helpful discussions.

\let\r\oldr

\bibliographystyle{plainnat}
\bibliography{cookflip}

\let\r\right


\appendix

\keep{%
\section{Proof of \texorpdfstring{$\lVert\cK_{\dt}\rVert < 1$}{|K\_dt| < 1}}
\label{sec:Kdtless1}

Recall the Fourier matrix representation \cref{eq:Kmn} of $\cK_{\dt}$:
\begin{align}
  {(K_{\dt})}_{mn}
  =
  \langle\cK_{\dt}\,\eigf_m,\eigf_n\rangle
  =
  \ee^{-\eigv_m^2\dt}\flipp_{mn}.
\end{align}
Take a test function $\psi(\z)$ with~$\lVert\psi\rVert=1$; then
\begin{align*}
  \Bigl\lvert
  \sum_{m,n\le N}
  {(K_{\dt})}_{mn}\,\psi_m\psi_n
  \Bigr\rvert^2
  &=
  \Bigl\lvert
  \sum_{m,n\le N}
  \ee^{-\eigv_m^2\dt}\,\psi_m\flipp_{mn}\,\psi_n
  \Bigr\rvert^2
  \\
  &\le
  \sum_{m \le N}
  \ee^{-2\eigv_m^2\dt}\l\lvert\psi_m\r\rvert^2
  \sum_{m',n\le N}\l\lvert\flipp_{m'n}\,\psi_n\r\rvert^2 \\
  &\le
  \ee^{-2\eigv_1^2\dt}\sum_{m \le N}
  \l\lvert\psi_m\r\rvert^2
  \sum_{m',n\le N}\l\lvert\flipp_{m'n}\,\psi_n\r\rvert^2\,.
\end{align*}
Taking the limit $N \rightarrow \infty$, we obtain
\begin{equation}
  \l\lVert\cK_{\dt}\psi \r\rVert^2
  \le
  \ee^{-2\eigv_1^2\dt} \lVert\psi\rVert^2 \lVert\cF\psi\rVert^2
  =
  \ee^{-2\eigv_1^2\dt} < 1,
\end{equation}
where we used~$\lVert\cF\psi\rVert=1$.
}

\section{Boundary layer solution for symmetric problem}
\label{sec:blayer}

\mathnotation{\ff}{\U}
\mathnotation{\Z}{Z}
\mathnotation{\kk}{k}

\begin{figure}
  \begin{center}
    \includegraphics[width=.6\textwidth]{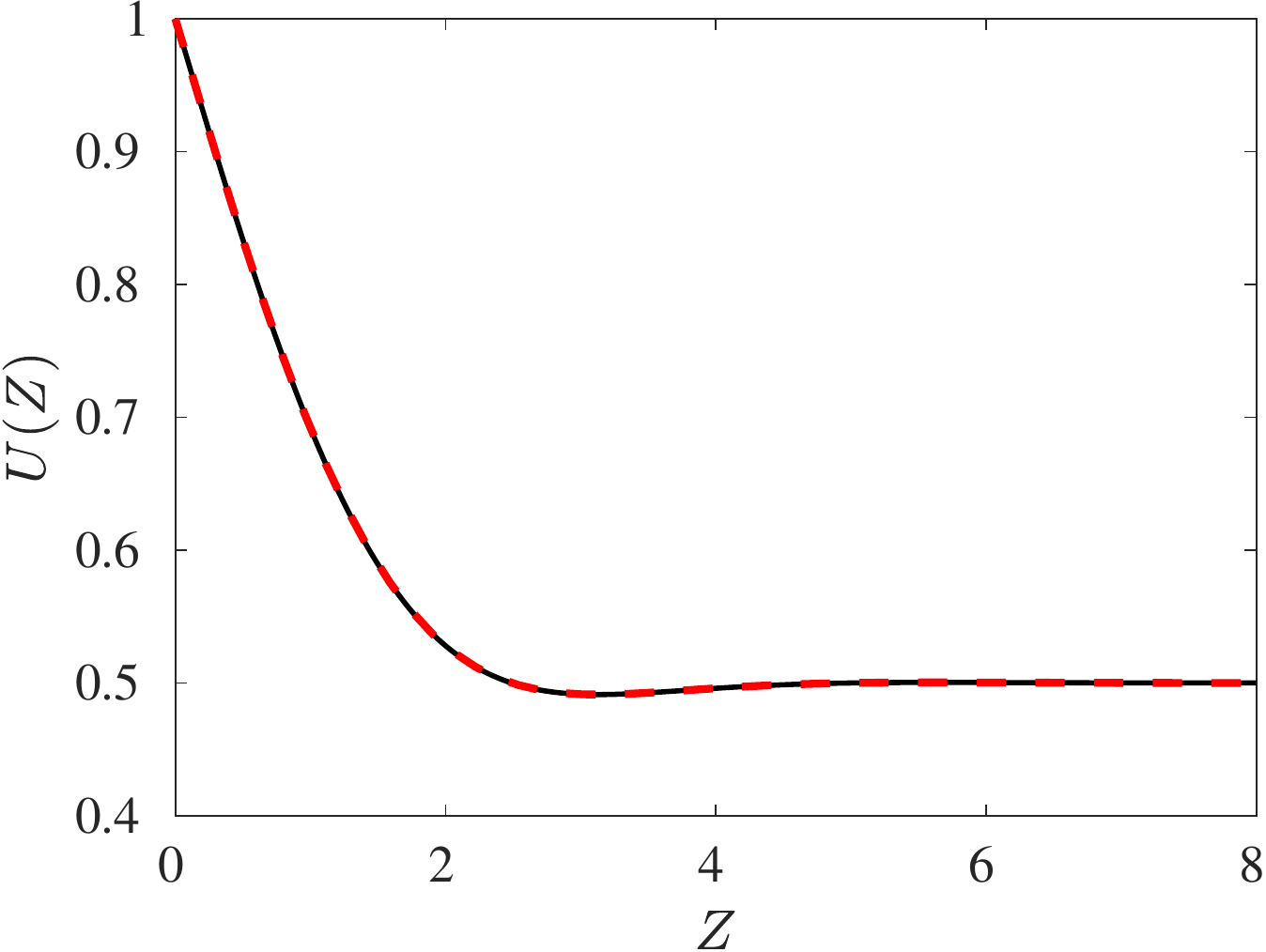}
  \end{center}
  \caption{Boundary layer solution from numerical evaluation
    of \cref{eq:intkk_muchbetter} (solid) and from the sum
    \cref{eq:Usym_again} for~$\dt=0.001$ (dashed).}
  \label{fig:flipheatbl}
\end{figure}

The fixed-point profile in the symmetric case is given by \cref{eq:Usym}:
\begin{equation}
  \U_{\dt}(\z)
  =
  \tfrac12
  +
  \sum_{\text{$m$ even}}
  \tanh(\eigv_m^2\dt/2)
  \,\hTs_m\eigf_m(\z).
  \label{eq:Usym_again}
\end{equation}
In this \nameCref{sec:blayer} we find the boundary layer structure
of~$\U_{\dt}(\z)$ in the rapid-flipping limit $\dt\rightarrow0$, for the case
of perfect conductors $\h_0=\h_1=\infty$, for which
\begin{equation}
  \eigv_m = m\pi\,,
  \qquad
  \eigf_m(\z) = \sqrt2\sin m\pi\z\,,
  \qquad
  \hTs_m = \frac{\sqrt2}{\pi m}\,.
\end{equation}
Define the stretched variables~$\Z = \z/\sqrt{\dt}$
and~$\kk_m = \eigv_m\sqrt{\dt}$; \cref{eq:Usym_again} is then
\begin{equation}
  \U_{\dt}(\Z\dt)
  =
  \tfrac12 +
  \sum_{\text{$m$ even}}
  \tanh(\kk_m^2/2)\, \frac{\sin(\kk_m\Z)}{\pi\kk_m}\,\Delta\kk_m
  \label{eq:sumkm}
\end{equation}
where~$\Delta\kk_m = \kk_{m+2} - \kk_{m} = 2\pi\sqrt{\dt}$.  In the limit
$\dt\rightarrow0$ the sum \eqref{eq:sumkm} becomes the integral
\begin{equation}
  \lim_{\dt\rightarrow0}  \U_{\dt}(\Z\dt)
  \rdef
  \ff(\Z)
  =
  \tfrac12 +
  \int_0^\infty
  \tanh(\kk^2/2)\, \frac{\sin(\kk\Z)}{\pi\kk}\dint\kk\,.
  \label{eq:intkk}
\end{equation}
This is convergent at~$\kk=0$ and~$\kk=\infty$, but for large~$\kk$ it only
converges because the fast oscillations of~$\sin(\kk\Z)$ overcome the slow
decay~$1/\kk$.  To improve convergence, first note that
\begin{equation}
  \int_0^\infty \frac{\sin(\kk\Z)}{\pi\kk} \dint\kk
  =
  \frac{1}{\pi}\int_0^\infty \frac{\sin x}{x} \dint x = \tfrac12\,,
  \qquad
  \Z>0.
\end{equation}
We then add and subtract~$1$ in the integrand of \cref{eq:intkk} to get
\begin{align}
  \ff(\Z)
  &=
  1
  +
  \int_0^\infty \l(\tanh(\kk^2/2)-1\r) \frac{\sin(\kk\Z)}{\pi\kk}\dint\kk
  \nonumber\\
  &=
  1
  -
  \frac{2}{\pi}
  \int_0^\infty \frac{1}{1 + \ee^{\kk^2}} \frac{\sin(\kk\Z)}{\kk}\dint\kk\,.
  \label{eq:intkk_muchbetter}
\end{align}
Now the integrand converges exponentially for large~$\kk$, making numerical
integration much easier.  \Cref{fig:flipheatbl} shows essentially perfect
agreement between \cref{eq:intkk_muchbetter,eq:Usym_again} for~$\dt=0.001$.
\matlabfootnote{The \MATLAB\ function \matlabcommand{flipheatbl} computes the
  boundary layer solution in \cref{fig:flipheatbl}.}

It's possible to evaluate the integral in~\eqref{eq:intkk_muchbetter} using
contour integration to obtain
\begin{align}
  \ff(\Z)
  =
  \frac{1}{2} +
  \sum_{n>0\ \text{odd}}
  \frac{2}{\pi n}\,
  \ee^{-\Z\sqrt{n\pi/2}}
  \sin\bigl(\Z\sqrt{n\pi/2}\bigr).
  \label{eq:intkk_muchbetter2}
\end{align}
However, the sum in~\eqref{eq:intkk_muchbetter2} is very poorly convergent for
small~$\Z$, and so numerical integration of~\eqref{eq:intkk_muchbetter} is
actually easier in that case.  Keeping one term in the sum, we can easily
show that the overshoot in the boundary layer solution is
near~$\Z=\sqrt{25\pi/8} \approx 3.13329$ (exact value~$3.12875\dots$).



\end{document}